\pdfoutput=1
\documentclass{article}
\usepackage{caption}
\usepackage{subcaption}
\usepackage{graphicx}
\usepackage{amsmath}
\usepackage{geometry}
\usepackage{authblk}
\usepackage{hyperref}
\usepackage[T1]{fontenc}
\usepackage[utf8]{inputenc}
\usepackage{float}
\usepackage{csquotes}

\usepackage{tikz}
	
\usetikzlibrary{backgrounds,fit,decorations.pathreplacing}

\title{A Universal Quantum Circuit Design for Periodical Functions}
\author{Junxu Li and Sabre Kais\thanks{Email: kais@purdue.edu}}
\affil{Department of Chemistry, Department of Physics and Astronomy, and

Purdue Quantum Science and Engineering Institute

Purdue University, West Lafayette, IN 47907, United States}

\begin{document}
\tikzset{meter/.append style={draw, inner sep=10, rectangle, font=\vphantom{A}, minimum width=30, line width=.8,
 path picture={\draw[black] ([shift={(.1,.3)}]path picture bounding box.south west) to[bend left=50] ([shift={(-.1,.3)}]path picture bounding box.south east);\draw[black,-latex] ([shift={(0,.1)}]path picture bounding box.south) -- ([shift={(.3,-.1)}]path picture bounding box.north);}}}
\tikzset{
 cross/.style={path picture={ 
\draw[thick,black](path picture bounding box.north) -- (path picture bounding box.south) (path picture bounding box.west) -- (path picture bounding box.east);
}},
crossx/.style={path picture={ 
\draw[thick,black,inner sep=0pt]
(path picture bounding box.south east) -- (path picture bounding box.north west) (path picture bounding box.south west) -- (path picture bounding box.north east);
}},
circlewc/.style={draw,circle,cross,minimum width=0.3 cm},
}
\maketitle
	
\begin{abstract}
We propose a universal quantum circuit design that can estimate any  arbitrary one-dimensional  periodic functions based on the corresponding Fourier expansion.
The quantum circuit contains N-qubits to store the information on the  different N-Fourier components and $M+2$ auxiliary qubits with $M = \lceil{\log_2{N}}\rceil$ for control operations. The  desired  output  will be measured in the last qubit $q_N$ with a time complexity of the computation of $O(N^2\lceil \log_2N\rceil^2)$.
We illustrate the approach by constructing the quantum circuit for the square wave function with accurate results obtained by direct simulations using the IBM-QASM simulator. The approach is general and can be applied to any arbitrary periodic function.

\end{abstract}

\section*{Introduction}
\label{sec_intro}

The field of quantum information and quantum computing advances in both software and hardware in the past few years.  The achievement of 72-qubit quantum chip, Sycamore, with programmable superconducting processor\cite{arute2019quantum} heralded a remarkable triumph towards quantum supremacy experiment \cite{preskill2018quantum}. On the other hand, the photonic quantum computer, Jiuzhang\cite{zhong2020quantum}, demonstrated  quantum computational advantages with Boson sampling using photons.
The blooming of hardware development by IBM, Google, IonQ and many others provokes tremendous enthusiasm developing quantum algorithms utilizing near term quantum devices and pursuit of application in various fields of science and engineering.
Recently there arises a growing body of research focusing on  quantum optimization\cite{moll2018quantum, zhou2020quantum}, solving linear system of  equations\cite{pan2014experimental, huang2019near, wossnig2018quantum}, electronic structure calculations \cite{huang2006entanglement, kandala2017hardware, xia2017electronic, parrish2019quantum, bian2021quantum, xia2020hybrid, xia2020qubit}, quantum encryption\cite{hu2020quantum, li2021practical}, variational quantum eigensolver (VQE)\cite{peruzzo2014variational, mcclean2016theory} for various problems\cite{sagastizabal2019experimental, ryabinkin2018constrained, nakanishi2019subspace} and open quantum dynamics\cite{barreiro2011open, hu2020quantumB, smart2021quantum, head2021capturing, pauls2013quantum} .
Recently, quantum machine learning further explored and implemented quantum software that could show advantages compared with the corresponding classical ones\cite{xia2018quantum, biamonte2017quantum, schuld2019quantum, huggins2019towards, dixit2021training, roy2021enhancement, sajjan2021quantum, wilson2021machine}.

However, difficulties arise inevitably when attempting to include nonlinear functions into quantum circuits.  For example, the  very existence of nonpolynomial activation functions guarantees that multilayer feedforward networks can approximate any functions\cite{leshno1993multilayer}. Even though, the nonlinear activation functions do not immediately correspond to the mathematical framework of quantum theory, which describes system evolution with linear operations and probabilistic observation. Conventionally, it is found extremely difficult generating these nonlinearities with a simple quantum circuit.  The alternative approach is to make a compromise, imitating the nonlinear functions with repeated measurements\cite{peruvs2000neural, zak1998quantum, cao2017quantum}, or with assistance of the Quantum Fourier Transformation\cite{schuld2015simulating} (QFT\cite{coppersmith2002approximate,weinstein2001implementation}). How to  simulate an  arbitrary function, especially nonlinear functions from a quantum circuit is an important issue to be addressed. 

In this paper, we proposed a universal design of quantum circuit, which is able to generate arbitrary finite continuous periodic 1-D functions, even nonlinear ones such as the square wave function, with the given Fourier expansion.
The output information is all stored in the last qubit, which could be measured for the function estimation, or  used as an intermediate state for following computation, as the case of estimating the nonlinear activation  function between layers in quantum machine learning methods.
We presented the details of the quantum circuit design in the first section followed by numerical simulation of the circuit imitating the square wave function on IBM-QASM. The final section contains complexity analysis and  further applications.

\section{Design of the quantum circuit}
\label{Design of the quantum circuit}

\begin{figure}[ht]
    \centering
    \begin{subfigure}[t]{0.35\textwidth}
        \centering
        \includegraphics[width=\textwidth]{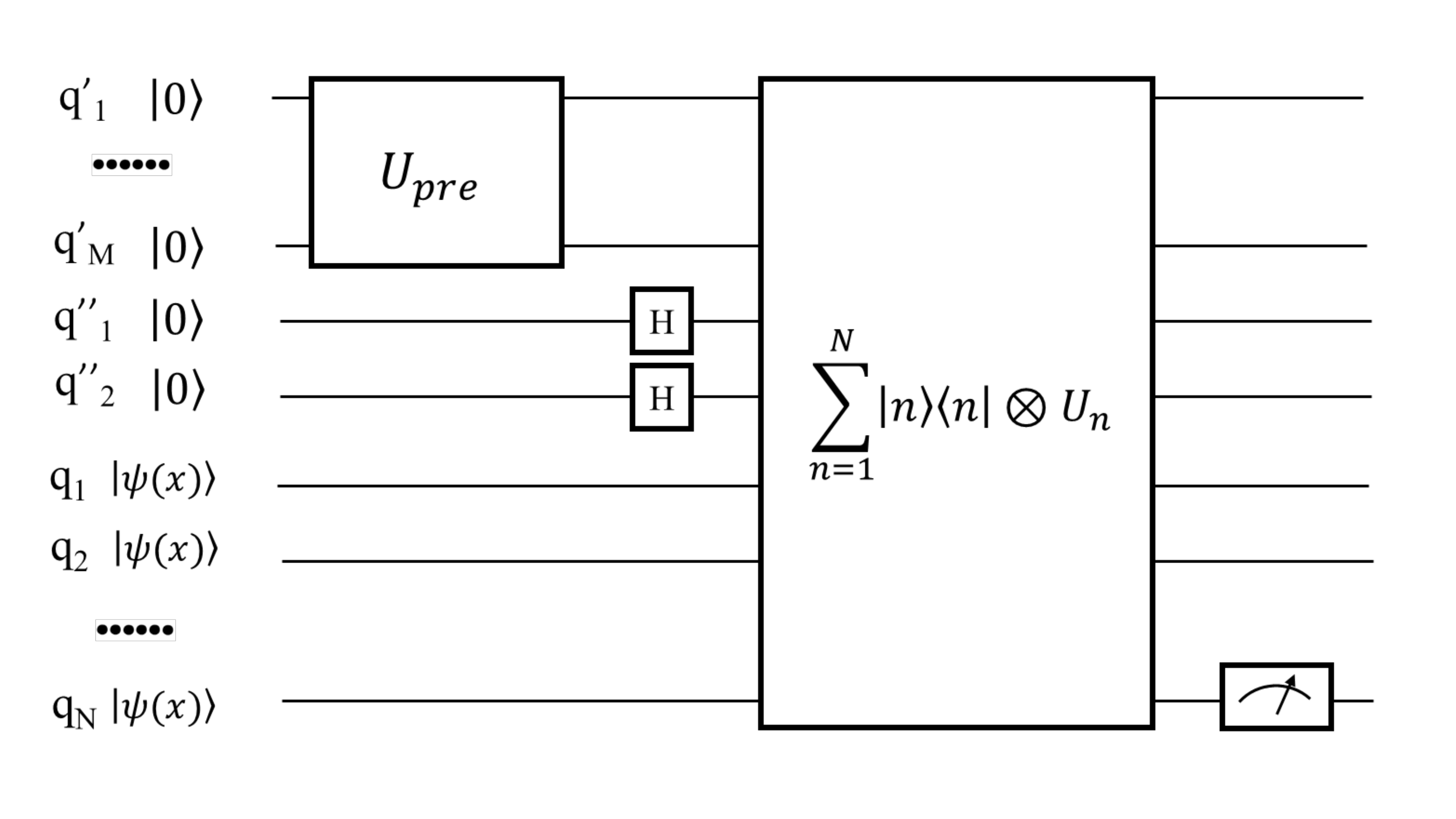}
        \caption{}
        \label{circuit_overview}
    \end{subfigure}
    \begin{subfigure}[t]{0.55\textwidth}
        \centering
        \includegraphics[width=\textwidth]{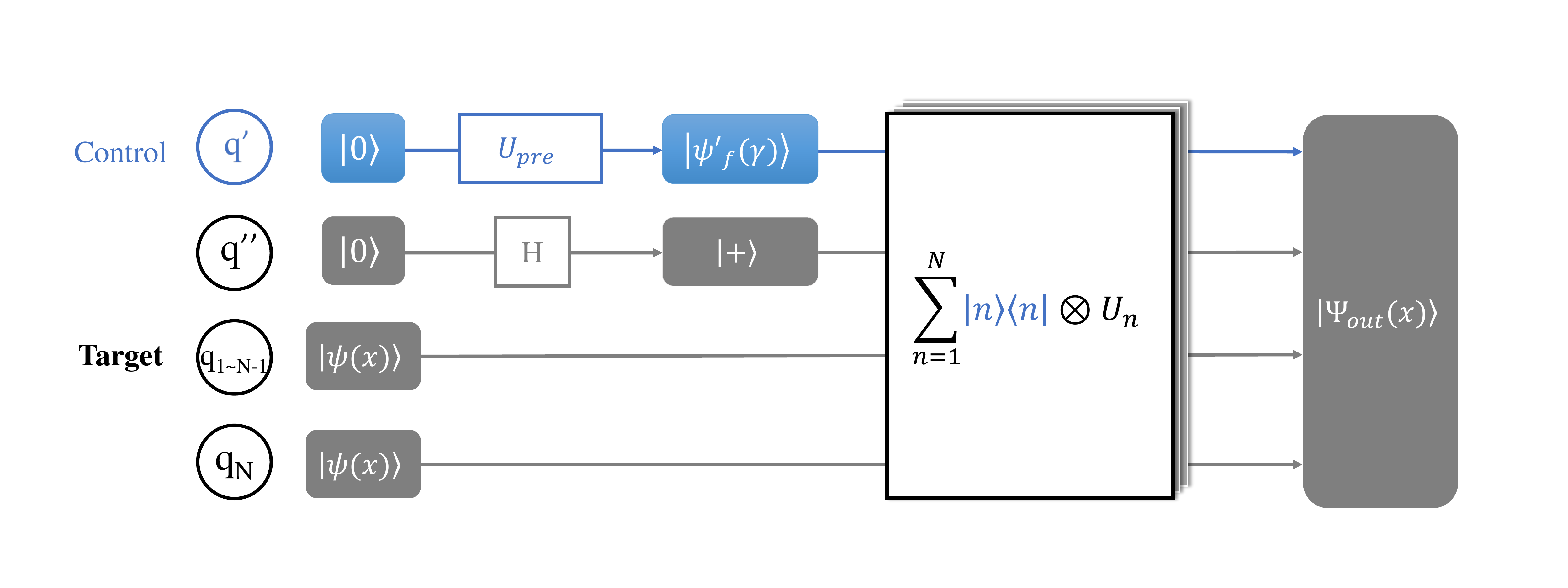}
        \caption{}
        \label{circuit_flowchart}
    \end{subfigure}
    ~
    \begin{subfigure}[t]{0.75\textwidth}
        \centering
        \includegraphics[width=\textwidth]{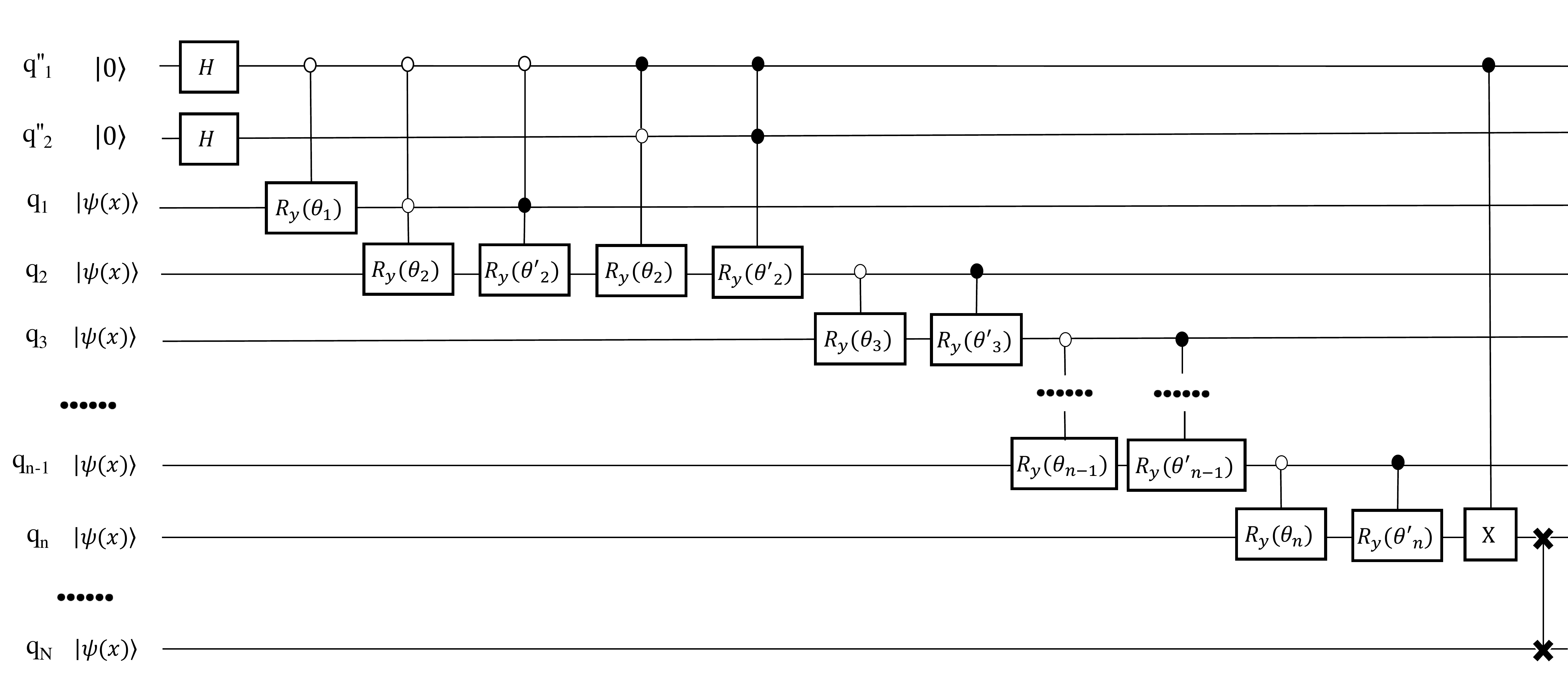}
        \caption{}
        \label{fig_un}
    \end{subfigure}
    \caption{
    {\bf Sketch of the quantum circuit estimating $F_N(x).$}
    \\
    Structure of the whole circuit is shown in fig.(\ref{circuit_overview}).
    There are two main modules.
    The first one contains $U_{pre}$ acting on the auxiliary qubits $q'$, and Hadamard gates acting on $q''$.
    The succeeding module is formed by $N$ controlled unitary operations denoted as $U_n$.
    The corresponding evolution is demonstrated in fig.(\ref{circuit_flowchart}),
    where $q'$ (blue color) are control qubits.
    $q'$ are converted to state $|\psi_f'(\gamma)\rangle$ under the operation $U_{pre}$,
    where $\gamma$ is determined by $F_N$.
    The detailed sketch of $U_n$ is shown in fig.(\ref{fig_un}).
    Initially, Hadamard gates are applied on $q''$, and a rotation $Y$ gate is applied on the first qubit $q_1$.
    All these three qubits  then acts as control qubits, while $q_2$ is the target in following operation.
    Next, qubits  $q_2, q_3,\cdots,q_n$ are connected as a chain with simple control rotation $Y$ gates.
    Finally a swap gate between $q_n$ and $q_N$ is included, ensuring that all the necessary information are stored in the last qubit $q_N$.
    }
    \label{fig_circuit}
\end{figure}

Consider a periodic 1-D function $F_N(x)$ that can be expanded as Fourier series with  $N$ nontrivial components,
\begin{equation}
    F_N(x) = 
    \sum_{n = 1}^{N}
    {
    a_n\cos(\frac{2\pi}{T}nx + b_n)
    }
    \label{FN}
\end{equation}
where $T$ is the period, and for simplicity we set $T = \pi$.
To construct the  quantum circuit that estimates the output function  $F_N(x)$, we need 
$N$-qubits to store  the input information, all of which are initially prepared at the  state $|\psi(x)\rangle = \cos x|0\rangle+\sin x|1\rangle$.
Additionally, there are $M + 2$ auxiliary qubits,  with  $M = \lceil{\log_2{N}}\rceil$ qubits assigned  $q'_1, \cdots, q'_M$ and the other two  are $q''_1, q''_2$.
All the  auxiliary qubits are initially  set as $|0\rangle$ states. 
Thus, the input state can be written as
\begin{equation}
    |\Psi_{in}(x)\rangle
    =
    |0\rangle^{\otimes M}_{q'}\otimes
    |0\rangle^{\otimes 2}_{q''}\otimes
    |\psi(x)\rangle^{\otimes N}_{q}
\end{equation}
where subscripts indicate the group of the three different registers.
Fig.(\ref{circuit_overview}) illustrates the structure of the quantum circuit design for the output function $F_N(x)$, while the detailed  evolution is demonstrated in fig.(\ref{circuit_flowchart}).  There are two main modules in the quantum circuit: The first one contains $U_{pre}$ acting on the auxiliary qubits $q'$, converting them from $|0\rangle$ to state $|\psi'_f\rangle$, and Hadamard gates acting on $q''$, converting them to states $|+\rangle=1/\sqrt{2}[|0\rangle+|1\rangle]$.  The intermediate state $|\psi'_f\rangle$ can be described as 
\begin{equation}
    |\psi'_f({\bf \gamma})\rangle =
    \sum_{n=0}^{2^M-1}
    \sqrt{\gamma_n}
    |n\rangle
    \label{eq_psif}
\end{equation}
where $\sum_{n=0}^{2^M-1}\gamma_n=1$ and $\gamma_n\geq0$. Details about the design of $U_{pre}$ and ${\bf \gamma_n}$ can be found in the supplementary materials.
The succeeding module is formed by $N$ controlled unitary operations, where $q'$ are the  control qubits for the target qubits  $q''$  and  $q$.  Denote the unitary operations as $U_n$,  with a general structure  shown in fig.(\ref{fig_un}). Initially, the Hadamard gates are applied on $q''$, and a rotation $Y$  gate is applied on the first qubit  $q_1$. All these three qubits then acts as control qubits, while $q_2$ is the target in following operation. Next, qubit $q_2, q_3,\cdots,q_n$ are connected as a chain with simple control rotation $Y$ gates. Finally a swap gate between $q_n$ and $q_N$ is included, ensuring that all the necessary information are stored in the last qubit $q_N$.

For simplicity, in the operation $U_n$, we define $w_{0,1}^k$ and $v_{0,1}^k$ as
\begin{equation}
\begin{split}
    {R_y(\theta_k)}^\dagger|0\rangle = \cos(w_0^k)|0\rangle + \sin(w_0^k)|1\rangle
    \\
    {R_y(\theta_k)}^\dagger|1\rangle = \cos(w_1^k)|0\rangle + \sin(w_1^k)|1\rangle
    \\
    {R_y(\theta'_k)}^\dagger|0\rangle = \cos(v_0^k)|0\rangle + \sin(v_0^k)|1\rangle
    \\
    {R_y(\theta'_k)}^\dagger|1\rangle = \cos(v_1^k)|0\rangle + \sin(v_1^k)|1\rangle
\end{split}
\label{vw}
\end{equation}
and $\alpha,\beta$ when $k\geq2$ as
\begin{equation}
    \alpha = \frac{1}{2}
    \prod_{k=2}^{n}
    |\sin(v_1^k - w_1^k)|
\end{equation}
\begin{equation}
    \beta_k = \arctan2(\sin2v_1^{k+1} - \sin2w_1^{k+1}, \cos2v_1^{k+1} - \cos2w_1^{k+1})
\end{equation}
while we have $\alpha = 1/2$ and $\beta = 2w_1^1$ when $k=1$.

To estimate $F_N(x)$, we need to ensure that
\begin{equation}
    C\sum_{n=1}^Na_n\cos(2nx+b_n)+\frac{1}{2}
    =
    \sum_{m=1}^{N}
    \gamma_m
    \alpha^m
    \prod_{k=1}^{m}\cos(2x-\beta_k^m)+\frac{1}{2}
    \label{constraint}
\end{equation}
where $C$ is a nonzero constant ensuring that $|CF_N(x)|\leq\frac{1}{2}$, and the superscript $m$ of $\alpha^m$, $\beta_k^m$ indicating that they belong to the operation $U_m$.
The right hand side of eq.(\ref{constraint}) is the probability to get the outcome result $|1\rangle$ when measuring $q_N$ itself after running the whole quantum circuit (More details can be found in the supplementary materials).

Thus, subsequent to the whole operation, the output state will be
\begin{equation}
    \begin{split}
    |\Psi_{out}(x)\rangle
    =
    \sqrt{-CF_N(x)+\frac{1}{2}}|\phi_0\rangle_{q',q'',q_1,\cdots,q_{N-1}}
    \otimes|0\rangle_{q_N}
    +\sqrt{CF_N(x)+\frac{1}{2}}|\phi_1\rangle_{q',q'',q_1,\cdots,q_{N-1}}
    \otimes|1\rangle_{q_N}
    \end{split}
    \label{output}
\end{equation}
where $|\phi_{0, 1}\rangle$ describing output state of all qubits and auxiliary qubits  except the last one $q_N$. Hence, information stored in $q_N$ is essential and sufficient.
After measuring $q_N$, the probability of getting $|1\rangle$ is an estimation of $F_N(x)$.  Whereas, it is also appropriate to apply succeeding operations on $q_N$, regarding it as an intermediate state for further computation.
The same quantum circuit structure  works when the input is in  a superposition states, namely $|\Psi_{in}^s({\bf x})\rangle$ describing a vector ${\bf x}$ that contains $L+1$ components,
\begin{equation}
    |\Psi_{in}^s({\bf x})\rangle
    =
    \sum_{l=0}^{L}
    {
    c_l|\Psi_{in}(x_l)\rangle_{q',q'',q}
    \otimes
    |\Phi_l\rangle_{Q}
    }
    \label{eq_input_sup}
\end{equation}
where $q',q'',q$ are entangled with some other qubits, namely in $Q$.
The superscript $s$ represents superposition, subscripts $q',q'',q$ and $Q$ indicate the group of different qubits in different registers, and $\sum_{l=0}^L{|c_l|}^2=1$.
$|\Phi_l\rangle$ is a complete orthogonal set of the subspace expanded by $Q$, ensuring that $\langle\Phi_{l'}|\Phi_l\rangle=\delta_{l'l}$.
After the whole operation, the output state  is given by
\begin{equation*}
    |\Psi_{out}^s({\bf x})\rangle
    =
    \sum_{l=0}^{L}
    {
    c_l|\Psi_{out}(x_l)\rangle_{q',q'',q}
    \otimes
    |\Phi_l\rangle_{Q}
    }
\end{equation*}
If we only focus on the subspace expanded by $q_N$ and $Q$, 
\begin{equation}
    |\Psi_{out}^s({\bf x})\rangle_{q_N,Q}
    =
    \sum_{l=1}^{L}
    {
    c_l
    \left[
    \sqrt{-CF_N(x_l)+\frac{1}{2}}
    |0\rangle_{q_N}
    +\sqrt{CF_N(x_l)+\frac{1}{2}}
    |1\rangle_{q_N}
    \right]
    \otimes
    |\Phi_l\rangle_{Q}
    }
    \label{eq_superposition}
\end{equation}
Then if $q_N$ is measured, probability to get result $|1\rangle$ will be
$\frac{1}{2}+C\sum_{l=0}^L|c_l|^2F_N(x_l)\frac{1}{2}$.
This property leads to potential applications in quantum algorithm development,
for example, the design of nonlinear activation between layers in quantum machine learning.

\section{Implementation: Simulation for the square wave function}

In this section we will demonstrate the quantum  circuit design with a trivial example: Simulation of the square wave function.  Consider the following Fourier expansion for the square wave function $f(x) = sign(\sin2x)$, the sum of the first 7 terms is given by,
\begin{equation}
    F(x) = \sum_{n=1 \in  odd}^{7}\frac{1}{n}\cos
    \left[n\cdot 2x - \frac{\pi}{2}\right]
    \label{eq_square}
\end{equation}
where $F(x)$ is an odd function with  $a_{2,4,6}=0$.   $q_{1,\cdots,7}$ are required carrying out  the input information, all of which will be initially prepared in the  state $|\psi(x)\rangle$. Additionally, we need 5 auxiliary qubits, denoting them as $q'_{1,2}, q''_{1,2,3}$, and $a_n', b_n'$ satisfy
\begin{equation}
    C\sum_{n=1}^{j}a'_n\cos(2nx+b'_n)
    =
    \sum_{n=1}^Na_n\cos(2nx+b_n)
    -\sum_{m=j+1}^{N}
    \gamma_m
    \alpha^m
    \prod_{k=1}^{m}\cos(2x-\beta_k^m)
    \qquad
    1\leq j<N
\end{equation}
Then for $n>1$ we set
\begin{equation}
    \left\{
    \begin{array}{lr}
        \gamma_n = \frac{2^{n-1}Ca'_n}{\alpha^n} &
        \\
        \beta_k^n = -\frac{b'_n}{n}, & b'_n\neq0
        \\
        \beta_k^n = \frac{\pi}{3}\delta_{1,0}-\frac{\pi}{3(n-1)}, &b'_n=0
    \end{array}
    \right.
    \label{betagamma}
\end{equation}

\begin{figure}[ht]
    \centering
    \begin{subfigure}[t]{0.85\textwidth}
        \centering
        \includegraphics[width=\textwidth]{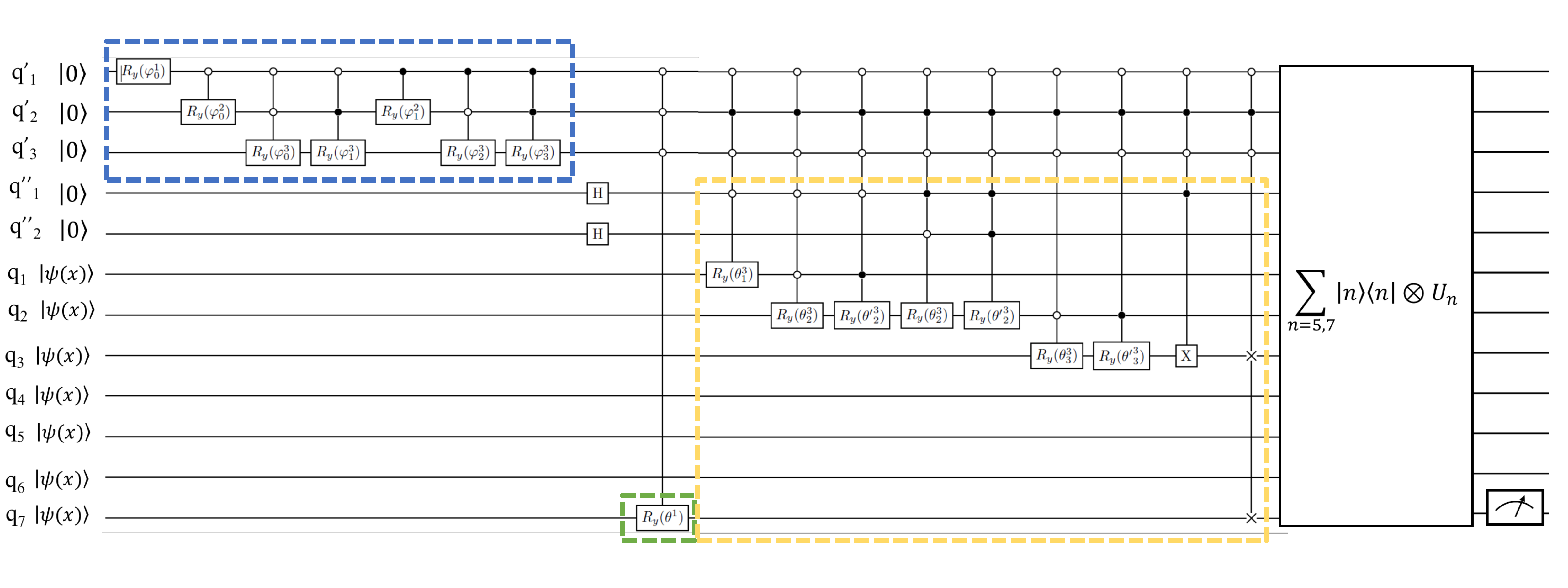}
        \caption{}
        \label{fig_detail}
    \end{subfigure}
    \centering
    \begin{subfigure}[t]{0.32\textwidth}
        \centering
        \includegraphics[width=\textwidth]{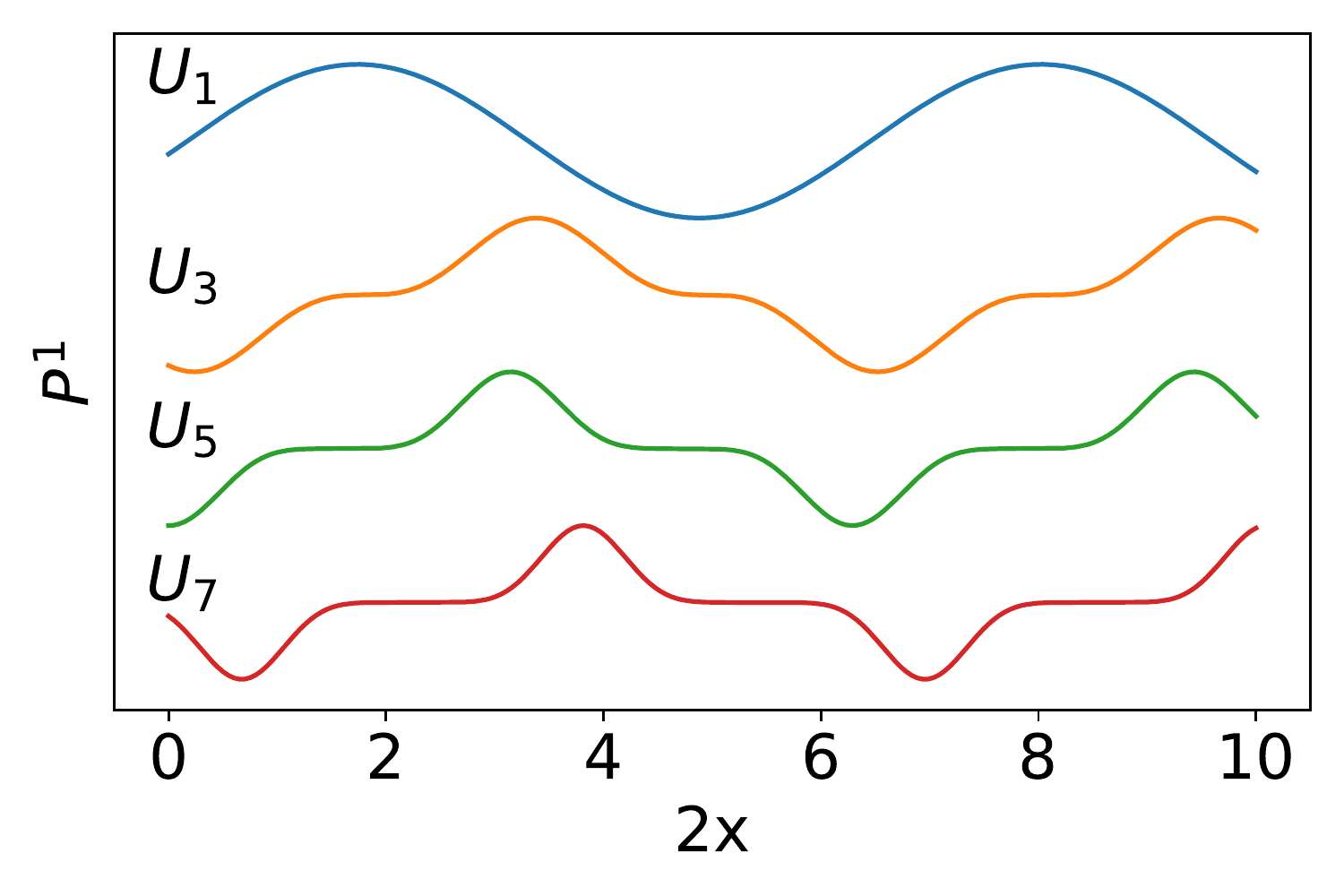}
        \caption{}
        \label{fig_part}
    \end{subfigure}
    \centering
    \begin{subfigure}[t]{0.32\textwidth}
        \centering
        \includegraphics[width=\textwidth]{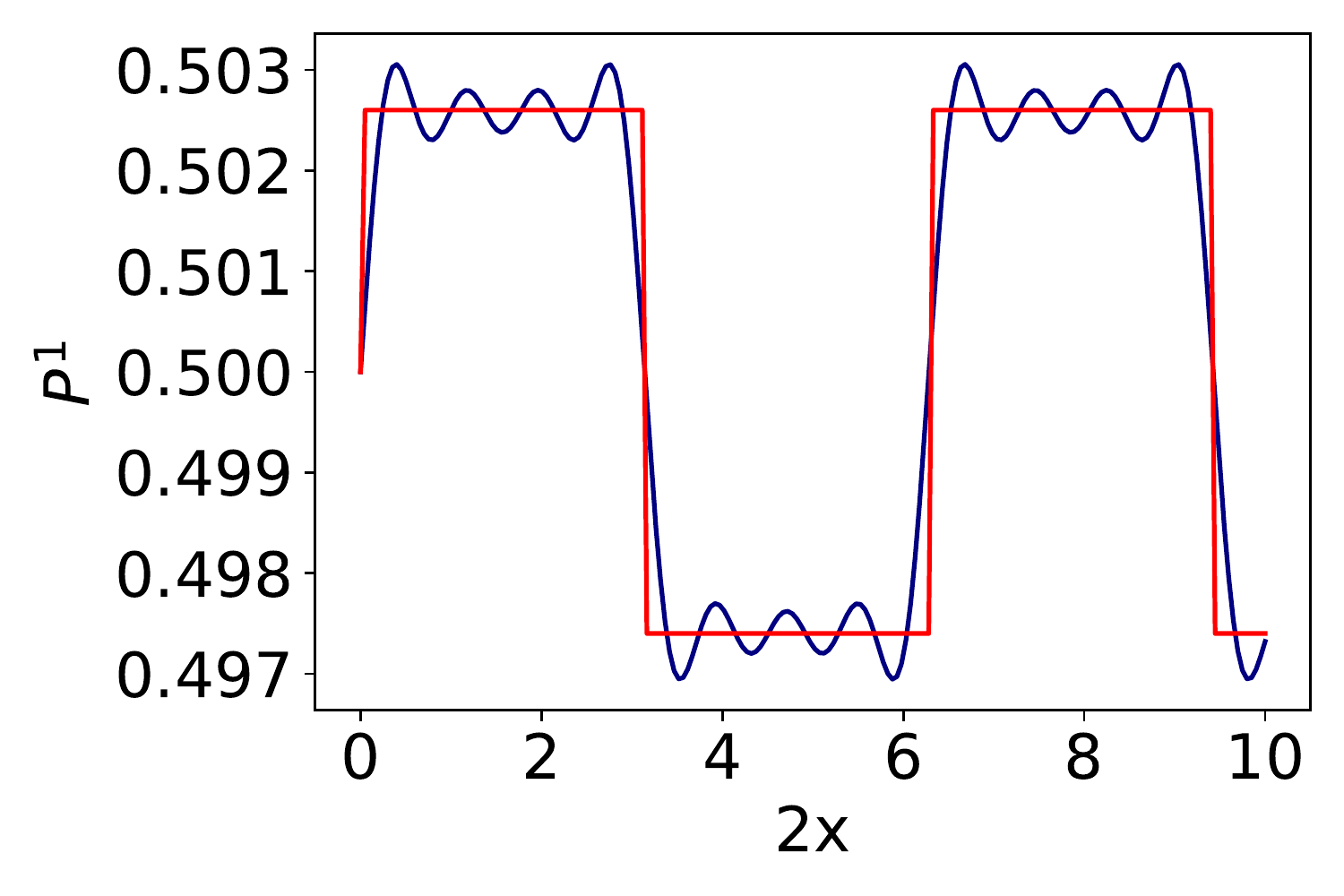}
        \caption{}
        \label{fig_square}
    \end{subfigure}
    \centering
    \begin{subfigure}[t]{0.23\textwidth}
        \centering
        \includegraphics[width=\textwidth]{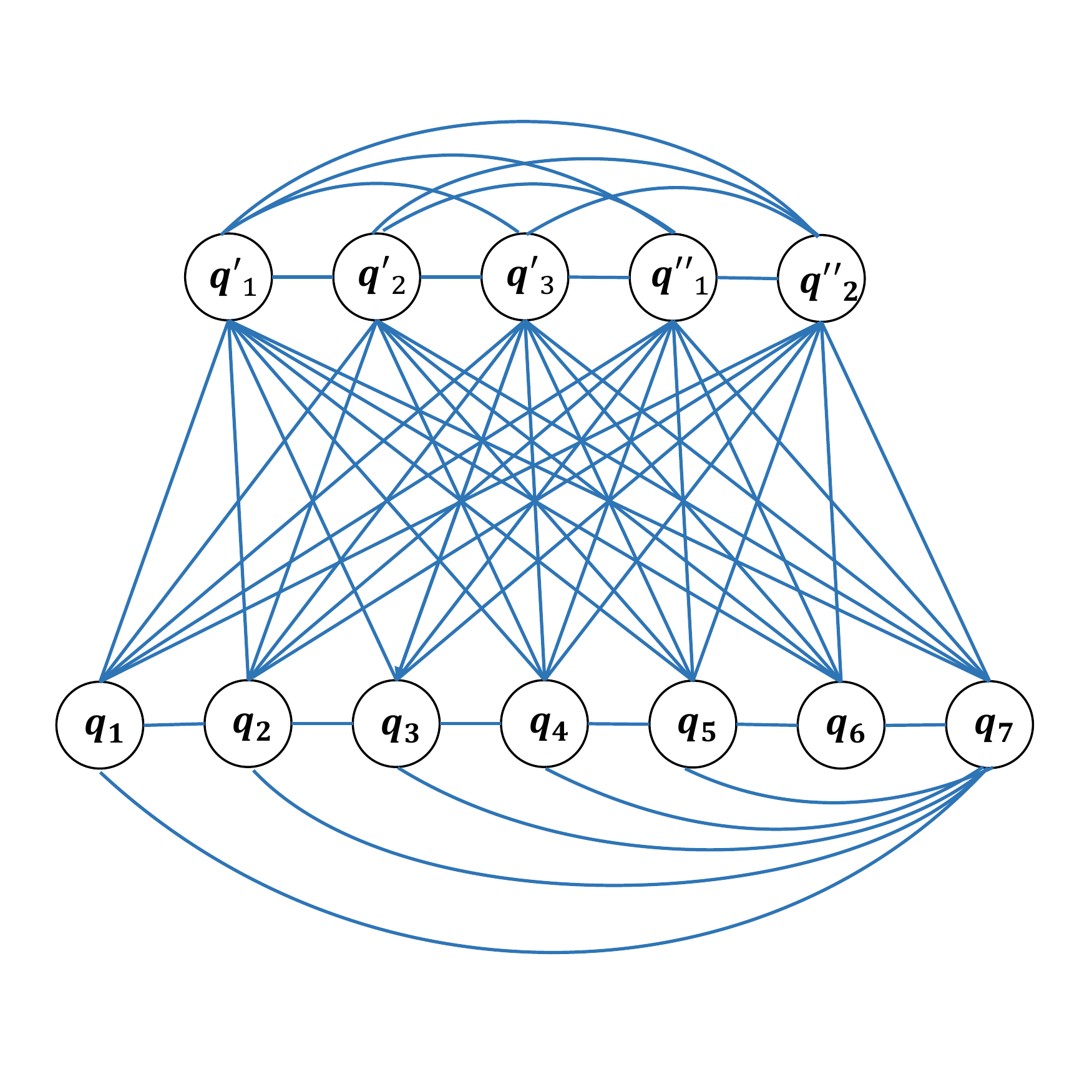}
        \caption{}
        \label{fig_connectivity}
    \end{subfigure}
    \caption{
    {\bf The quantum circuit estimating 1-D square wave function.}
    \\
    Fig.(\ref{fig_detail}) is a sketch of the quantum circuit estimating square wave function.
    Operations in the blue block is $U_{pre}$, which converts $q'$ into state $|\psi'_f(\gamma)\rangle$ in eq.(\ref{eq_psif}).
    Due to the fact that $\gamma_{2,4,6}$ are all zero, $U_{2,4,6}$ disappears in the circuit.
    As shown in the green block, $U_1$ is a single rotation $Y$  gate acts on the last qubit $q_7$ directly.
    The other $U_n$ acts on $q''_{1,2}$, qubits $q_{1, 2, \cdots, j}$ and $q_7$.
    Attributable to the space, here we only plot details of $U_3$, as shown in the yellow block.
    Numerical simulation results are included in fig.(\ref{fig_part},\ref{fig_square}).
    $P_1$ is probability to get result $|1\rangle$ when measuring the last qubit $q_7$ after the whole operation.
    $x$ is the variable in the input state $|\psi(x)\rangle$.
    Contributions of each single operation $U_j$ alone is shown in fig.(\ref{fig_part}).
    Amplitudes are not included when plotting fig.(\ref{fig_part}).
    In fig.(\ref{fig_square}), the blue curve represents the sum of contributions of all $U_{1,3,5,7}$, which is as well the expected result when measuring $q_7$.
    Meanwhile, the red curve is the original shape of square wave functions.
    Fig.(\ref{fig_connectivity}) is a sketch of the connectivity structure.
    Each node represents a single qubit. 
    If two qubits are connected via a blue curve, then there is at least one 2-qubit gate acting on them.
    }
    \label{fig_simulation}
\end{figure}

Eq.(\ref{betagamma}) ensures that $\gamma_{2,4,6}$ are all zero, so that we do not need to construct  $U_{2,4,6}$ in the circuit.   We need to stress the fact  that the above setting is not optimal, especially when one prefer a greater value of $|C|$ instead of a shallow circuit.  Fig.(\ref{fig_detail}) is a scheme of the whole operation.   Operations in the blue block is $U_{pre}$, which converts $q'$ into state $|\psi'_f(\gamma)\rangle$ in eq.(\ref{eq_psif}).   Due to the fact that $\gamma_{2,4,6}$ are all zero, then there is no need for construct   $U_{2,4,6}$  in the quantum  circuit.
As shown in the green block, $U_1$ is a single rotation $Y$  gate acts on the last qubit  $q_7$ directly. The other $U_n$ acts on $q''_{1,2}$,  $q_{1, 2, \cdots, j}$ and $q_7$.  For illustration, we only plot details of $U_3$, as shown in the yellow block.   Based on eq.(\ref{constraint}), eq.(\ref{eq_square}) can be rewritten as
\begin{equation*}
\begin{split}
    F(x) = &-0.8211 \cdot \cos(2x-4.8812)
    + 18.9339 \cdot \cos^3(2x-0.2384)
    \\
    &- 15.6030 \cdot \cos^5(2x-0.0046) 
    - 9.1429 \cdot \cos^7(2x-0.6732)
\end{split}
\end{equation*}
Contributions of each single operation $U_j$ alone is shown in fig.(\ref{fig_part}). The sum of all their contributions are shown in fig.(\ref{fig_square}), which is  the output result. $P_1$ is probability to get the outcome result $|1\rangle$ when measuring the last qubit $q_7$ after the whole operation. Fig.(\ref{fig_connectivity}) is a sketch of the connectivity structure.
Each node represents a single qubit.  If two qubits are connected via a blue curve, then there is at least one 2-qubit gate acting on them.  All auxiliary qubits $q'$ and $q''$ are connected with each other. Meanwhile, all qubits  $q$ are connected to all of the other auxiliary qubits.
For any $1\leq n<N$, there is also a connection between $q_n$ and its neighbor $q_{n+1}$.  The last qubit  $q_N$ is connected to all other qubits.

\label{simulation}

\begin{figure}[ht]
    \centering
    \begin{subfigure}[t]{0.32\textwidth}
        \centering
        \includegraphics[width=\textwidth]{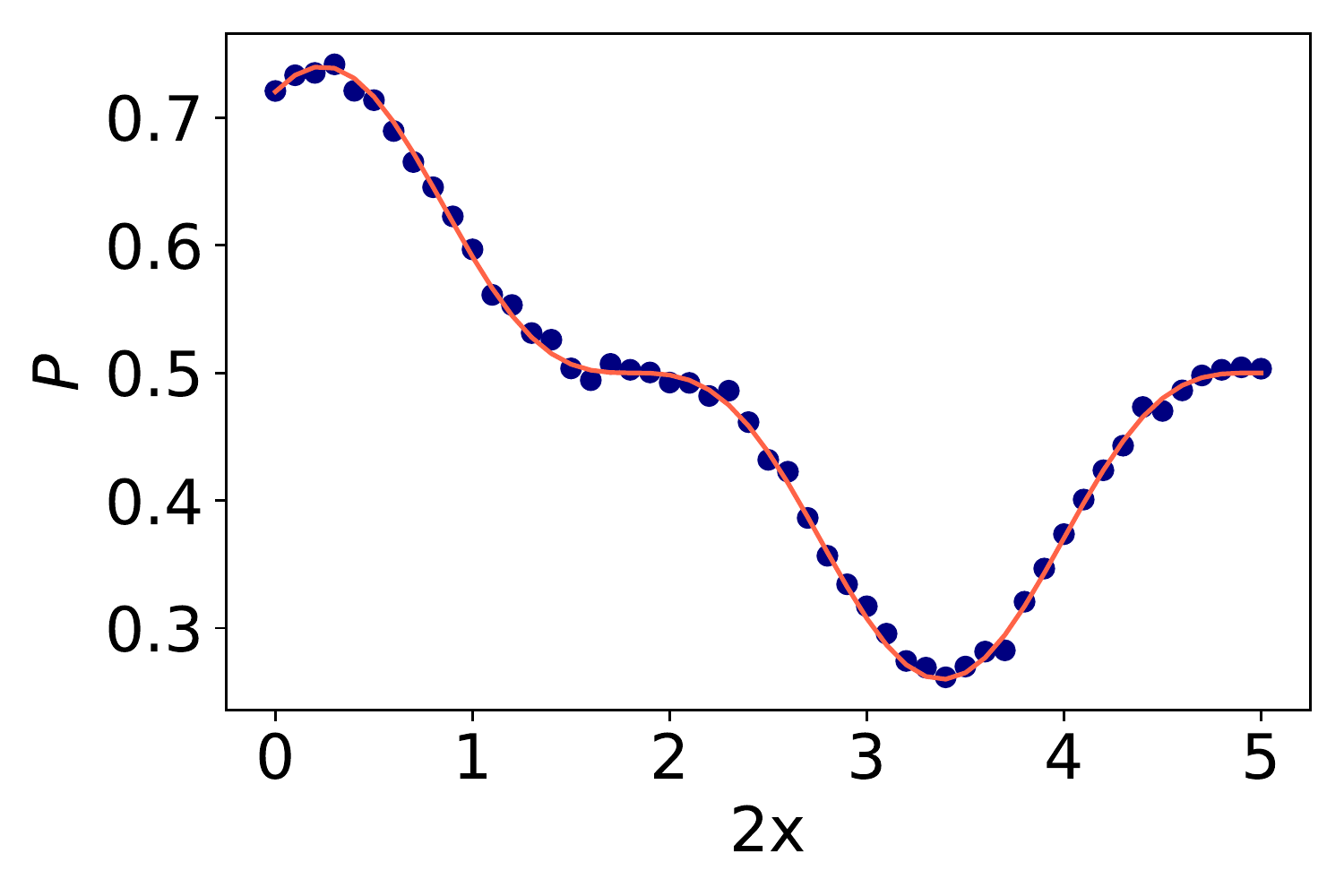}
        \caption{}
        \label{fig_u3qsam}
    \end{subfigure}
    \centering
    \begin{subfigure}[t]{0.32\textwidth}
        \centering
        \includegraphics[width=\textwidth]{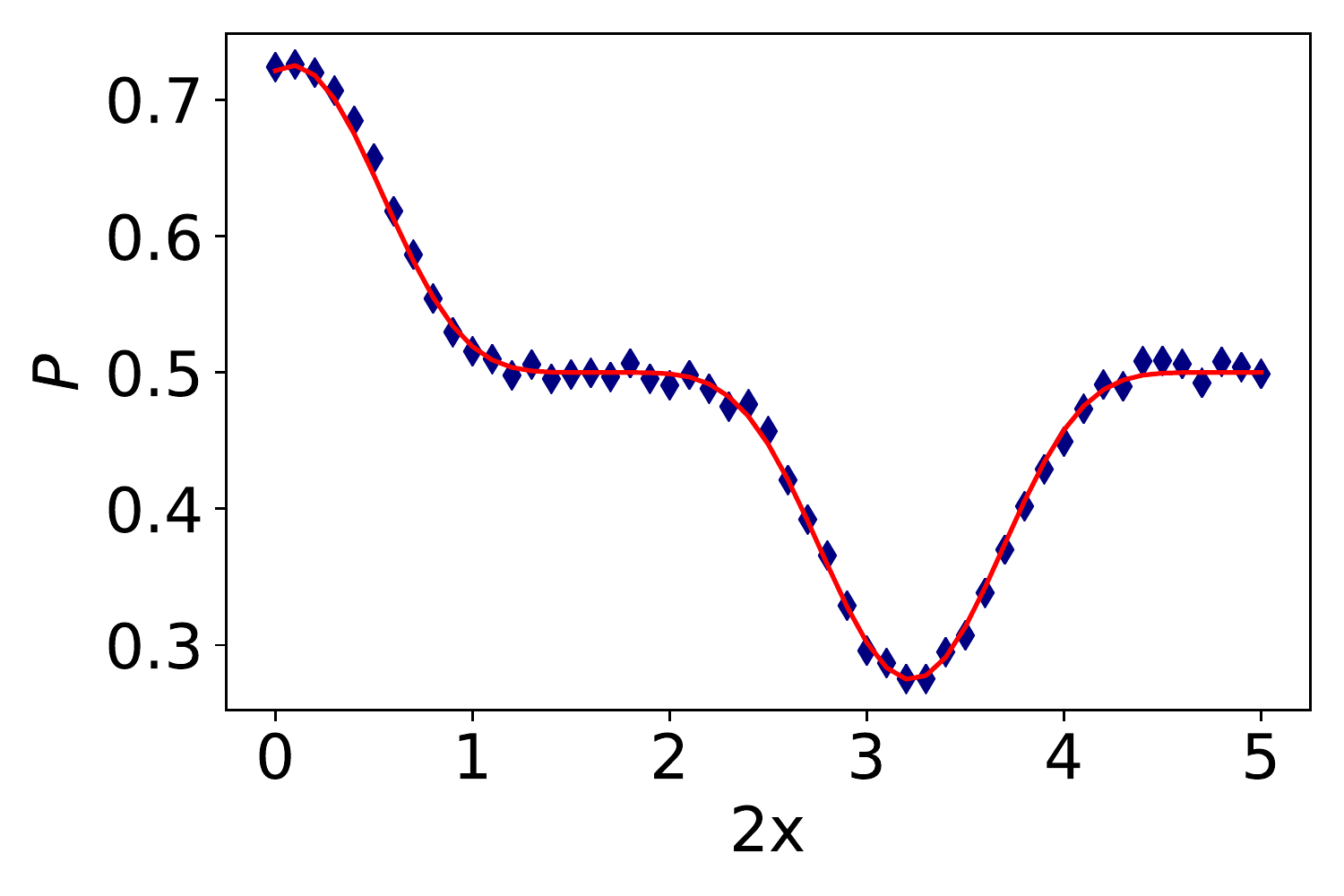}
        \caption{}
        \label{fig_u5qsam}
    \end{subfigure}
    \centering
    \begin{subfigure}[t]{0.32\textwidth}
        \centering
        \includegraphics[width=\textwidth]{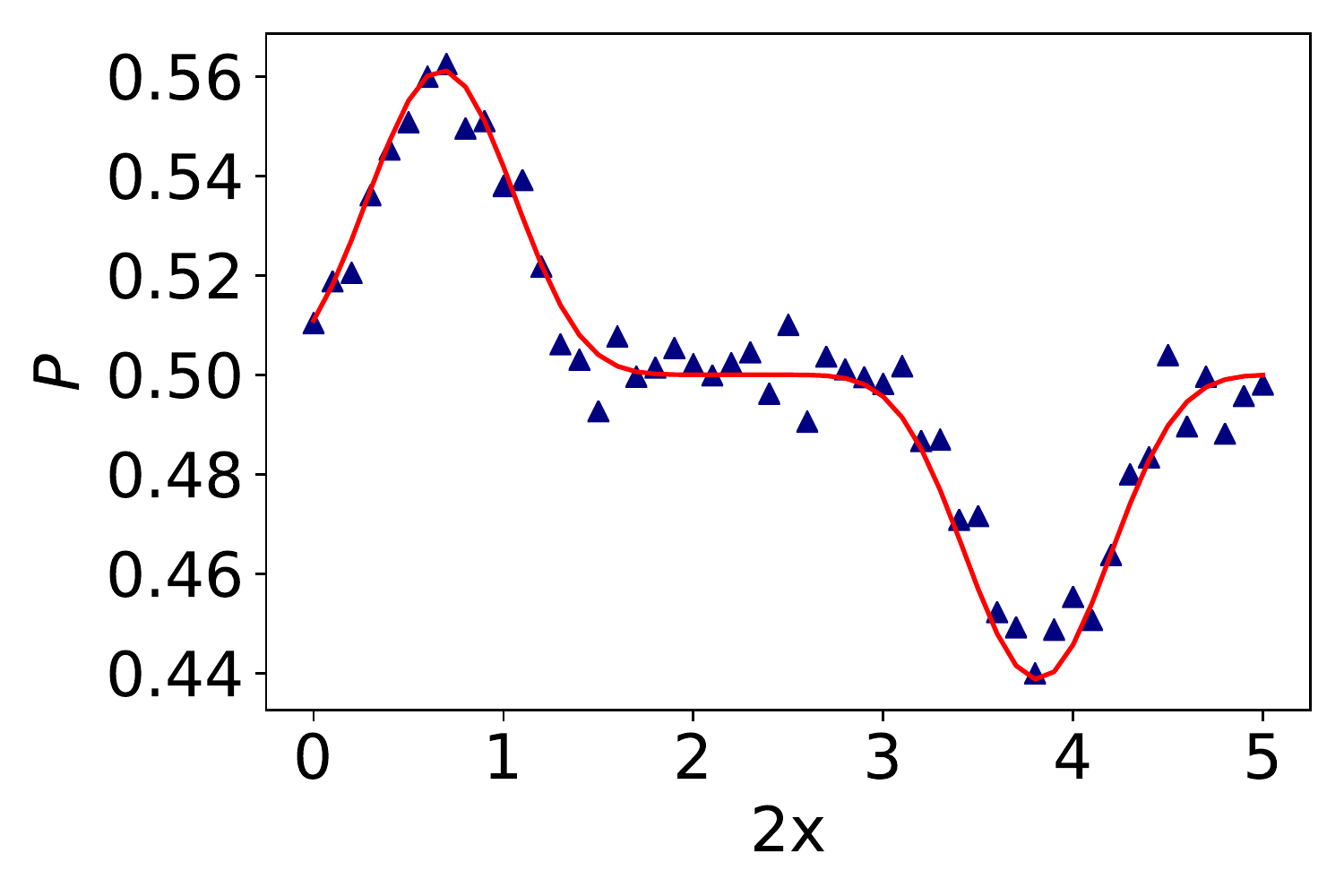}
        \caption{}
        \label{fig_u7qsam}
    \end{subfigure}
    \caption{
    {\bf Implementation on IBM QASM simulator.}
    \\
    Fig.(\ref{fig_u3qsam}) is numerical simulation results of $U_3$, contributing component $\cos^3(2x-0.2384)$, while fig.(\ref{fig_u5qsam}) and fig.(\ref{fig_u7qsam}) corresponds to $U_5$ and $U_7$, calculating components $\cos^5(2x-0.0046)$ and $\cos^7(2x-0.6732)$.
    Red lines represent the theoretical prediction, and blue dots(or diamonds, triangles) represent the results on IBM QASM simulator.
    $P$ is the probability to get state $|0\rangle$ or $|1\rangle$ when measuring the last qubit. 
    (Generally, we prefer to get state $|1\rangle$ when calculating a positive component in the expansion, and $|0\rangle$ when calculating a negative one.)
    For each single dot we collect data from 8192 iterative measurements.
    }
    \label{fig_qsam}
\end{figure}

Further, we also implemented $U_{3,5,7}$ independently based on IBM QASM simulator, as shown in fig.(\ref{fig_qsam}), where for each single dot we collect data from 8192 iterative measurements. Results from IBM QASM simulator fit  well with  the theoretical prediction corresponding to $U_3$ and $U_5$, as shown in fig.(\ref{fig_u5qsam}) and fig.(\ref{fig_u7qsam}).  More details of the simulations can be found in the supplementary materials.

\section{Time complexity}
\label{sec_complexity}
We will compare the time complexity for  three various situations: In the first situation we consider classical inputs, while for others we consider some unknown quantum states as inputs.
One can either estimate $x$ from the input and calculate $F_N(x)$ classically, or apply the method presented in this article to estimate the outputs. 

{\bf Situation I.}
Estimate $F_N(x)$ with a classical input $x$:

To estimate a periodic function $F_N(x)$ within error $\epsilon$, $O(\frac{1}{\epsilon})$ times of measurement are required\cite{gilyen2019quantum}.
Initially $N$ $R_y$ rotation gates are required for mapping $x$ into quantum state $|\psi(x)\rangle^{\otimes N}$. 
Moreover, we need $M+2$ auxiliary qubits, where $M=\lceil \log_2N\rceil$ for control qubits. 
$U_{pre}$ contains $O(exp(M))$ multi control gates, and in each of them there are at most $M-1$ control qubits. The time scaling of n-control gates is  $O(n^2)$\cite{barenco1995elementary}, thus the time complexity of $U_{pre}$ is $O(N\lceil \log_2N\rceil^2)$.

Consider the basic unit $U_n$. 
When $n>2$, there are $4$ Control-Control rotation gates, $4n-8$ control-rotation gates and one optional 2-qubit swap gate. 
Thus, it takes time $O(n\lceil \log_2n\rceil^2)$ to implement the operation $|n\rangle\langle n|\otimes U_n$.
After taking the $M$ auxiliary qubits into account, $M$ control qubits are added to all gates in $U_n$.
Therefore, it takes time $O(n)$ to achieve a single $U_n$
Notice that there are $N$ similar $U_n$, where $n=1,2,\cdots, N$, time complexity to finish all $\sum_{n=1}^N|n\rangle\langle n|\otimes U_n$ is $O(N^2\lceil \log_2N\rceil^2)$.

Totally, to derive the estimation $CF_N(x)$ within error $\epsilon$, 
the time complexity is of order $O(N^2\lceil \log_2N\rceil^2/\epsilon)$, which is still polynomial.
On the other hand, time complexity to estimate $F_N(x)$ for a single $x$ based on Taylor expansion is also polynomial to $N$.
Hence under this situation there is no speedup comparing with the classical calculation.

{\bf Situation II.}
Estimate $F_N(x)$ with input quantum state $|\psi(x)\rangle^{\otimes N}$, where the value of $x$ is still unknown:

The only difference from the previous situation is that the mapping process now can be skipped. 
Still, time consuming is of order $O(N^2\lceil \log_2N\rceil^2/\epsilon)$.
By contrast, to calculate $F_N(x)$ from $|\psi(x)\rangle^{\otimes N}$ classically, the first step is to derive $x$ from the input states.
It requires $O(\frac{1}{\epsilon})$ measurements to get $\cos x$ within error $\epsilon$, and then the time complexity to estimate $F_N(x)$ is polynomial to $N$. Still, under this situation there is no speedup comparing with the classical calculation.

{\bf Situation III.}
Estimate $\sum_{l=0}^L|c_l|^2F_N(x_l)$ with input quantum state
$|\Psi_{in}^s({\bf x})\rangle$, as described in eq.(\ref{eq_input_sup}):

Here we denote $N'$ as the number of qubits that form state $|\Phi\rangle$, and $L=2^{N'}-1$.
Even though more variables are introduced into the input, we do not need to change anything in the quantum circuit. 
To get an estimation of $C\sum_{l=0}^L|c_l|^2F_N(x_l)$ with the quantum method, time consuming is still of order $O(N^2\lceil \log_2N\rceil^2/\epsilon)$, which does not depend on the scale of $N'$(Number of qubits in $Q$).
However, in the classical method, as $c_j$ are unknown initially, one must estimate them before calculating $\sum_{l=0}^L|c_l|^2F_N(x_j)$. 
Time complexity of quantum tomography is exponential in $N'$, or at least polynomial in $N'$ with shadow quantum tomography. 
The time complexity of quantum method is only determined by $N$, while the classical method is at least polynomial in $N'$\cite{aaronson2019shadow}.
Thus, when $N'>>N$, the quantum method will lead to polynomial speedup comparing with the classical one.

Consider the task to estimate $|c_0|^2F(x_0)+|c_1|^2F(x_1)$, where $|c_0|^2+|c_1|^2=1$.
For simplicity, here we set $F(x)=cos^3(x-0.2384)$, which can be implemented by $U_3$ itself, excluding the $q'$ registers.
(Structure of $U_3$ can be found in Fig.(2) in the main article and Fig.(S2) in the SM.)
In {\bf Situation III}, the input states $|\Psi_{in}^s({\bf x})\rangle$ can be described by Eq.(9).

\begin{figure}[htp]
    \centering
    \includegraphics[width=0.75\linewidth]{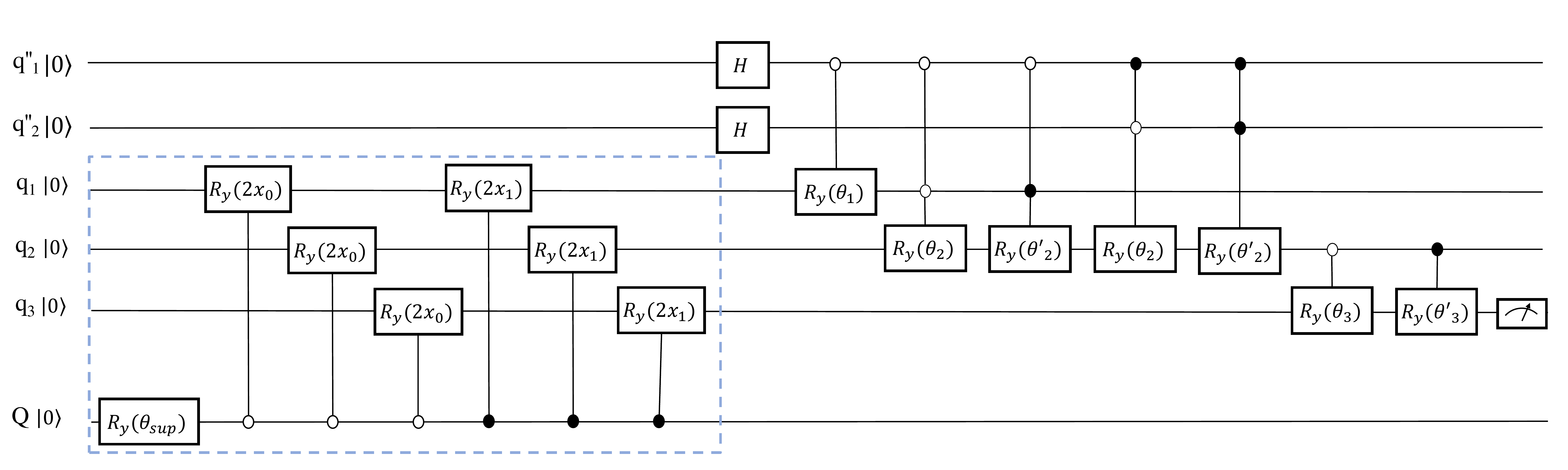}
    \caption{
    {\bf A sketch of the quantum circuit that calculates $|c_0|^2F(x_0)+|c_1|^2F(x_1)$.}
    \\
    There are totally 6 qubits,
    $q''_{1,2}$ and $q_{1,2,3}$ implementing $U_3$, and qubit $Q$ corresponding to Eq.(\ref{eq_input_sup}).
    Initially, all qubits are set as $|0\rangle$.
    Operations in the dashed blue square converts them into state
    $|\Psi_{in}^s(x_0, x_1)\rangle$, as shown in Eq.(\ref{eqs_input_sup}).
    Then, operation $U_3$ is applied on $q'$ and $q$, implementing the calculation of $F(x)$.
    After the whole operation, we only to measure $q_3$ itself to estimate $|c_0|^2F(x_0)+|c_1|^2F(x_1)$.
    $P_0$, the probability to get result $|0\rangle$ when measuring $q_3$ is presented in Eq.(\ref{eq_p0}).
    }
    \label{figs_u3sup}
\end{figure}

In the task estimating $|c_0|^2F(x_0)+|c_1|^2F(x_1)$, we only need one qubit as $Q$ preparing the initial states.
A sketch of the quantum circuit calculating $|c_0|^2F(x_0)+|c_1|^2F(x_1)$ is shown in Fig.(\ref{figs_u3sup}).
There are totally 6 qubits,
$q''_{1,2}$ and $q_{1,2,3}$ implementing $U_3$, and qubit $Q$ corresponding to Eq.(\ref{eq_input_sup}).
Initially, all qubits are set as $|0\rangle$.
Operations in the dashed blue square converts them into state
\begin{equation}
    |\Psi_{in}^s(x_0, x_1)\rangle
    =
    c_0|\Psi_{in}(x_0)\rangle_{q',q'',q}
    \otimes
    |0\rangle_{Q}
    +
    c_1|\Psi_{in}(x_1)\rangle_{q',q'',q}
    \otimes
    |1\rangle_{Q}
    \label{eqs_input_sup}
\end{equation}
where $|\Psi_{in}(x)\rangle_{q',q'',q}=|0\rangle^{\otimes2}_{q'}\otimes|\psi(x)\rangle^{\otimes3}_{q}$, and $|\psi(x)\rangle = \cos x|0\rangle+\sin x|1\rangle$.
Coefficients $c_{0,1}$ are determined by the gate $R_y(\theta_{sup})$ acting on $Q$, leading to
\begin{equation*}
    c_0 = \cos(\frac{\theta_{sup}}{2})
\end{equation*}
\begin{equation*}
    c_1 = \sin(\frac{\theta_{sup}}{2})
\end{equation*}
Then, operation $U_3$ is applied on $q'$ and $q$, implementing the calculation of $F(x)$.
After the whole operations, the output state  is given by
\begin{equation*}
    |\Psi_{out}^s({x_0, x_1})\rangle
    =
    c_0|\Psi_{out}(x_0)\rangle_{q',q'',q}
    \otimes
    |0\rangle_{Q}
    +
    c_1|\Psi_{out}(x_1)\rangle_{q',q'',q}
    \otimes
    |1\rangle_{Q}
\end{equation*}
where $|\Psi_{out}(x)\rangle$ is described by Eq.(8).

In this example, $P_0$, which represents the probability to get result $|0\rangle$ when measuring $q_3$, can be calculated as
\begin{equation}
    P_0=
    \cos^2(\frac{\theta_{sup}}{2})[0.2362\cos^3(2x_0-0.2384)+0.5]
    +
    \sin^2(\frac{\theta_{sup}}{2})[0.2362\cos^3(2x_1-0.2384)+0.5]
    \label{eq_p0}
\end{equation}
Thus, we only need to measure $q_3$ itself to estimate $|c_0|^2F(x_0)+|c_1|^2F(x_1)$.

\begin{figure}[htp]
    \centering
    \begin{subfigure}[t]{0.45\textwidth}
        \centering
        \includegraphics[width=\textwidth]{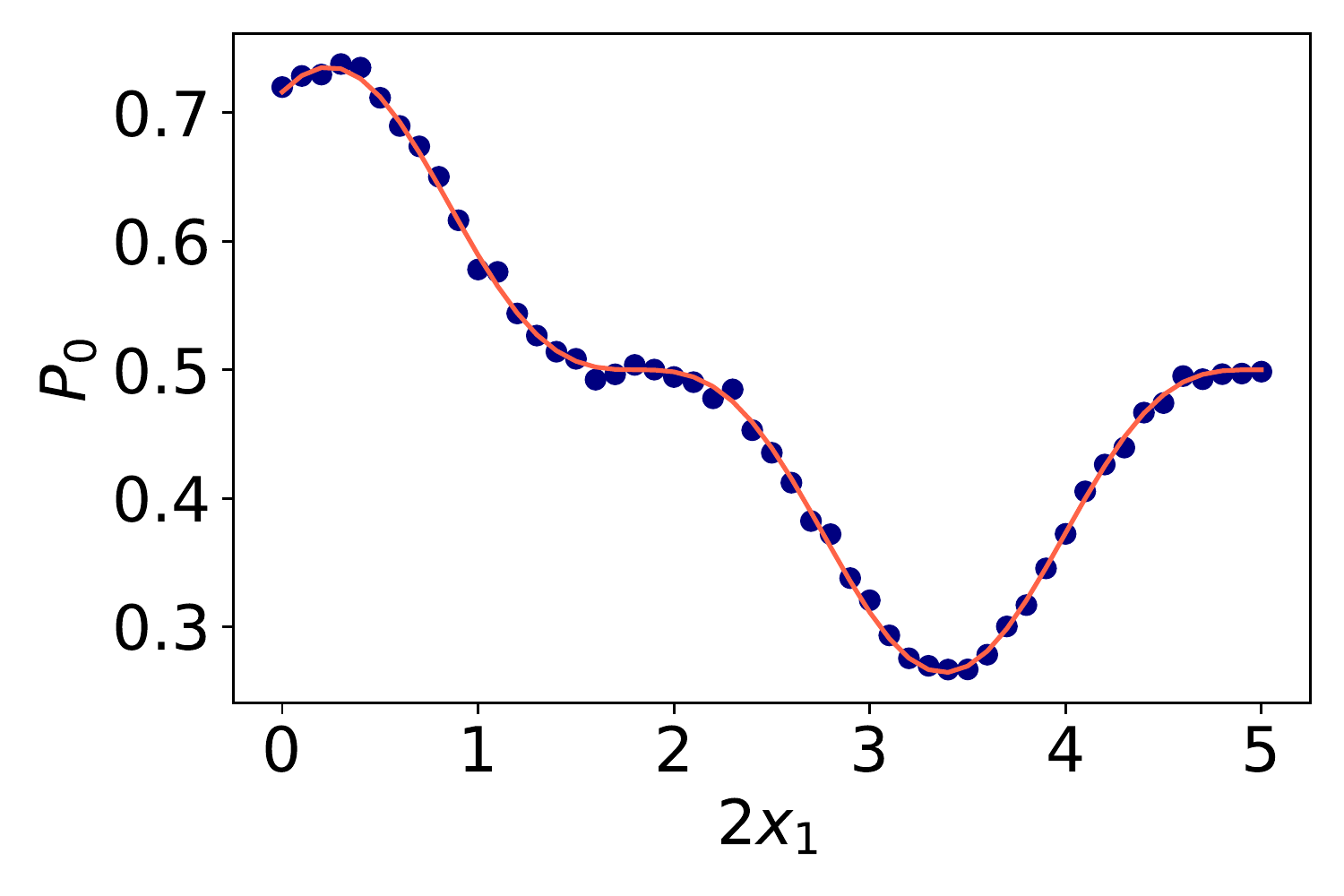}
        \caption{}
        \label{fig_origin}
    \end{subfigure}
    \centering
    \begin{subfigure}[t]{0.45\textwidth}
        \centering
        \includegraphics[width=\textwidth]{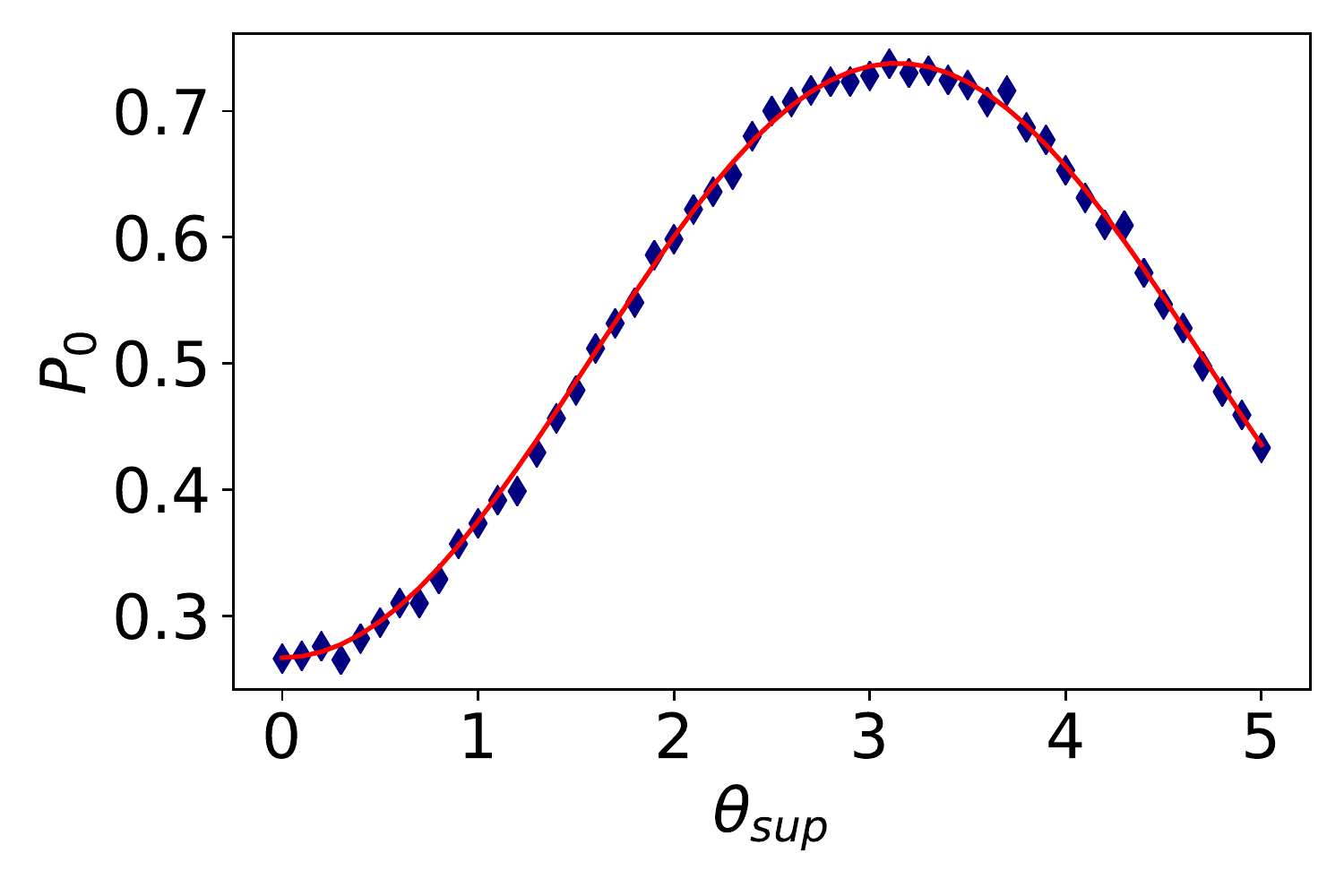}
        \caption{}
        \label{fig_sup}
    \end{subfigure}
    \caption{
    {\bf Implementation on IBM QASM simulator.}
    \\
    Here we present the simulation results obtained from IBM QASM simulator.
    In both figures $P_0$ indicates the probability to get result $|0\rangle$ when measuring $q_3$ after the whole operation shown in Fig.(\ref{figs_u3sup}).
    Blue dots or diamonds represent the simulation results from IBM QASM simulator,
    while the red curve is theoretical prediction based on Eq.(\eqref{eq_p0}).
    In Fig.(\ref{fig_origin}), we set $x_0=0$, $\theta_{sup}=\pi$, and collect $P_0$ under various $x_1$ values.
    $P_0$ shows the shape of $F(x_1)$, as we set $\theta_{sup}=\pi$ so that $x_0$ has no contribution to $P_0$.
    In Fig.(\ref{fig_sup}), we set $x_0=3.4$, $x_1=0.2$, and collect $P_0$ under various $\theta_{sup}$ values.
    Theoretically $P_0$ should have a single sin function shape, and the simulation results fit the prediction very well.
    For each single dot we collect data from 8192 iterative measurements.
    }
    \label{fig_sup_qsam}
\end{figure}

Further, we implement the operation shown in Fig.(\ref{figs_u3sup}) on IBM QASM simulator.
Simulation results obtained from IBM QASM simulator are shown in Fig.(\ref{fig_sup}),
where $P_0$ indicates the probability to get result $|0\rangle$ when measuring $q_3$ after the whole operation.
Blue dots or diamonds represent the simulation results from IBM QASM simulator,
while the red curve is theoretical prediction based on Eq.(\ref{eq_p0}).
In Fig.(\ref{fig_origin}), we set $x_0=0$, $\theta_{sup}=\pi$, and collect $P_0$ under various $x_1$ values.
$P_0$ shows the shape of $F(x_1)$, as we set $\theta_{sup}=\pi$ so that $x_0$ has no contribution to $P_0$.
In Fig.(\ref{fig_sup}), we set $x_0=3.4$, $x_1=0.2$, and collect $P_0$ under various $\theta_{sup}$ values.
Theoretically $P_0$ should have a single sin function shape, and the simulation results fit the prediction very well.
For each single dot we collect data from 8192 iterative measurements.

In the simulation only one qubit, $q_3$, is measured, and $|c_0|^2F(x_0)+|c_1|^2F(x_1)$ can be estimated from $P_0$, the probability to get result $|0\rangle$.
Therefore, with the quantum algorithm as we proposed, we do not need to estimate the exact values of $c_l$, so that the quantum method leads to speedup comparing with the classical version in {\bf situation III}, especially when $L$ is large.

\section{Conclusion}
\label{sec_conclusion}

 Here, we  propose a universal  quantum circuit design for any arbitrary one-dimensional periodic function.  The inputs are sufficient qubits prepared at the same state, while the last one will represent the output outcome. One can either estimate the exact value from repeating measurements on the last qubit, or regard it as an intermediate state prepared for succeeding operations. Superposition in the input leads to similarly superposition in the output, which leads to speedup under some certain circumstances, especially when dealing with unknown quantum inputs. As an simple example we illustrate the quantum circuit design for the  square wave function.  Both exact simulations and implementation on IBM-QASM gave very accurate result and illustrate the power of this proposed general design.   This general approach might be used to construct an appropriate  quantum circuit for the electronic wave function in periodic solids and  materials, moreover in quantum machine learning particularly simulating  the non-linear function used in the network.  

\section*{Acknowledgment}
\label{acknowledgment}
The authors would like to thank Zinxuan Hu and Blake Wilson for the helpful suggestions and discussions. 
We acknowledge the financial support in part by the National Science Foundation under award number 1955907 and funding by the U.S. Department of Energy (Office of Basic Energy Sciences) under Award No.de-sc0019215.

\bibliography{ref}
\end{document}


\beginsupplement
\tikzset{meter/.append style={draw, inner sep=10, rectangle, font=\vphantom{A}, minimum width=30, line width=.8,
 path picture={\draw[black] ([shift={(.1,.3)}]path picture bounding box.south west) to[bend left=50] ([shift={(-.1,.3)}]path picture bounding box.south east);\draw[black,-latex] ([shift={(0,.1)}]path picture bounding box.south) -- ([shift={(.3,-.1)}]path picture bounding box.north);}}}
\tikzset{
 cross/.style={path picture={ 
\draw[thick,black](path picture bounding box.north) -- (path picture bounding box.south) (path picture bounding box.west) -- (path picture bounding box.east);
}},
crossx/.style={path picture={ 
\draw[thick,black,inner sep=0pt]
(path picture bounding box.south east) -- (path picture bounding box.north west) (path picture bounding box.south west) -- (path picture bounding box.north east);
}},
circlewc/.style={draw,circle,cross,minimum width=0.3 cm},
}
\maketitle


\section{Preparations}
\label{sec_s1}
Consider a 1-D continuous bounded function $f(x)$ defined on $x\in[x_1, x_2]$, where $x_2 > x_1$, one can always expand $f(x)$ into a periodic function $F(x)$ in the whole 1-D space, where the cycle of $F(x)$ is $T = 2(x_2 - x_1)$
\begin{equation}
F(x) = 
\left\{
\begin{array}{lr}
     f(x), & x\in\left[x_1 + kT,\quad  x_2 + (k+1/2)T\right)\\
     f(x_2 - x), & x\in\left[x_1 + (k+1/2)T,\quad  x_2 + kT\right)\\
\end{array}
\right.
\quad 
k\in\mathrm{Z}
\label{eqs_fourier}
\end{equation}
As a continuous periodic function, $F(x)$ can be expanded with Fourier series
\begin{equation}
F(x) = \frac{1}{2} a_0 + 
\sum_{n = 1}^{\infty}
{
a_n\cos(\frac{2\pi}{T}nx + b_n)
}
\end{equation}
where $a_n$ and $b_n$ are defined as
\begin{equation}
    a_n =
    {\left\{
    \left[
    \int_Tdx {F(x)\sin(\frac{2\pi}{T} nx)}
    \right]^2
    +
    \left[
    \int_Tdx {F(x)\cos(\frac{2\pi}{T} nx)}
    \right]^2
    \right\}}^{\frac{1}{2}}
\end{equation}
\begin{equation}
    b_n =
    -\arctan2{
    \left[
    \int_Tdx {F(x)\sin(\frac{2\pi}{T} nx)},
    \int_Tdx {F(x)\cos(\frac{2\pi}{T} nx)}
    \right]
    }
\end{equation}
The first $N$ nontrivial components in Fourier expansion are given by
\begin{equation}
    F_N(x) = 
    \sum_{n = 1}^{N}
    {
    a_n\cos(\frac{2\pi}{T}nx + b_n)
    }
    \label{_FN}
\end{equation}
For simplicity, in the main article and following discussion we assume that $T=\pi$.
Hence, we have
\begin{equation}
    F_N(x) = 
    \sum_{n = 1}^{N}
    {
    a_n\cos(n\cdot 2x + b_n)
    }
    \label{FN}
\end{equation}
Given a certain $F_N$ as eq.(\ref{FN}), our goal is to design a quantum circuit that can estimate $F_N(x)$ with $N$ copies of the input state $|\psi(x)\rangle$, where
\begin{equation}
    |\psi(x)\rangle = \cos x|0\rangle+\sin x|1\rangle
\end{equation}
If more than $N$ nontrivial terms are obtained from eq.(S1), then $F_N(x)$ becomes an approximation of the original function $F(x)$.
We will focus on the construction of quantum circuit estimating $F_N(x)$,
while the remainder of the Fourier series (if any) is not covered.
Therefore, accuracy of the quantum algorithm is similar to the approximation $F_N(x)$, 
and in general to reach a higher accuracy more terms would be required.

\section{Design of $U_n$}
\label{sec_ucell}
\begin{figure}
    \centering
    \includegraphics[width=0.85\linewidth]{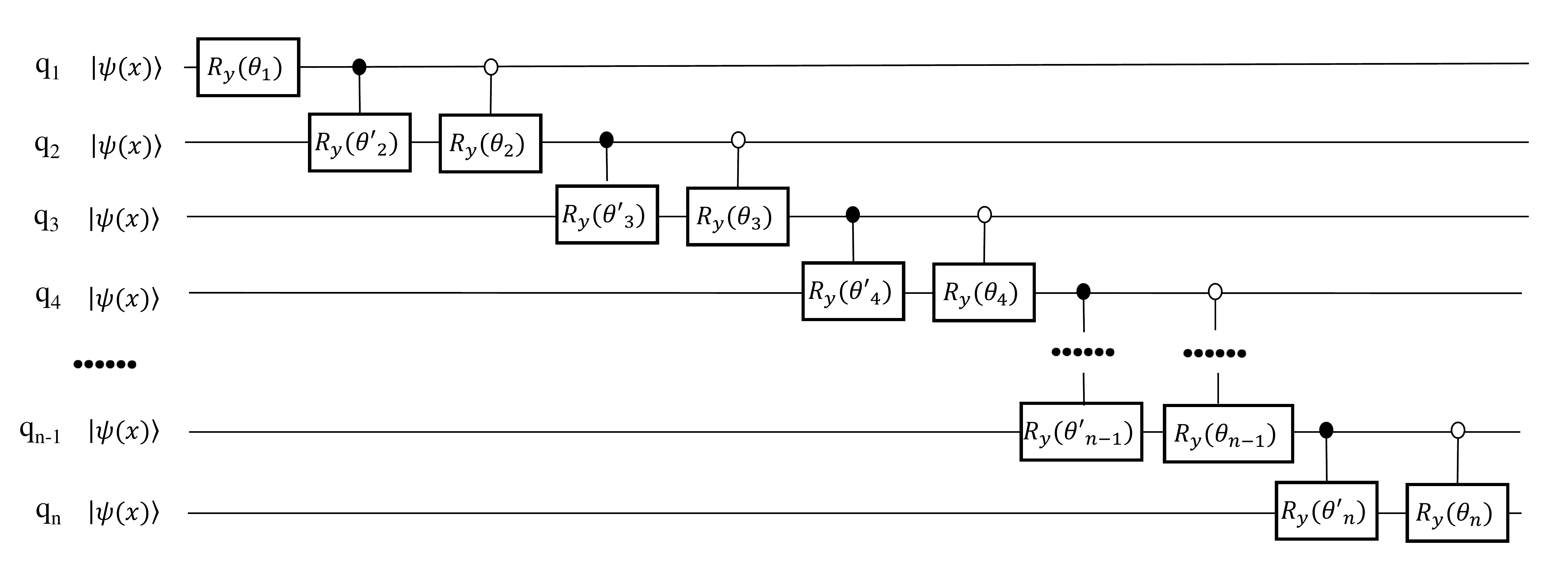}
    \caption{
    {\bf A sketch of a simple chain of qubits connected by control gates.}
    \\
    To convert the three qubits $q'_1,q'_2,q'_3$ into $|\psi'_f({\bf\gamma})\rangle$, the iterative process is repeated for three times.
    In the first iterative 'loop', we only need to apply a single rotation gate on $q'_1$.
    In the second loop, $q'_1$ performs as a control qubit.
    Two different rotation gates are applied on $q'_2$ corresponding to the two eigenstates of $q'_1$.
    Similarly, in the third loop, $q'_1$ and $q'_2$ together play the role of control qubits, and there are four rotation gates acting on $q'_3$.
    Totally, three are one rotation y gate acting on $q'_1$, two control rotation gates and four control control rotation gates.
    }
    \label{figs_chain}
\end{figure}

Consider the simple operation shown in fig.(\ref{figs_chain}), which is a chain of qubits connected by control gates.
All qubits are set as $|\psi(x)\rangle$ initially.

Denote the probability to get result $|0\rangle$ when applying a single Z measurement on qubit $q_n$ as $P_0^n$, and probability to get $|1\rangle$ as $P_1^n$, we have $P_0^n+P_1^n=1$.
We denote $U_n^{0,1}$ as rotation gates act on qubit $q_n$ when the control qubit $q_{n-1}$ is $|0\rangle$ or $|1\rangle$, with an exception that $U_1$ is a single qubit operation acts on $q_1$.
Corresponding to sketch fig.(\ref{figs_chain}), we have $U_0^n=R_y(\theta_n))$, and $U_1^n=R_y(\theta'_n)$.
As $F_N(x)$ only contains a single variable, here we set all control gates as control rotation-y gates. 
For simplicity, in the operation $U_n$, we define $w_{0,1}^k$ and $v_{0,1}^k$ as
\begin{equation}
\begin{split}
    {R_y(\theta_k)}^\dagger|0\rangle = \cos(w_0^k)|0\rangle + \sin(w_0^k)|1\rangle
    \\
    {R_y(\theta_k)}^\dagger|1\rangle = \cos(w_1^k)|0\rangle + \sin(w_1^k)|1\rangle
    \\
    {R_y(\theta'_k)}^\dagger|0\rangle = \cos(v_0^k)|0\rangle + \sin(v_0^k)|1\rangle
    \\
    {R_y(\theta'_k)}^\dagger|1\rangle = \cos(v_1^k)|0\rangle + \sin(v_1^k)|1\rangle
\end{split}
\label{vw}
\end{equation}
Due to there is no control qubits controlling the first qubit, we have
$v_0^1=w_0^1$, $v_1^1=w_1^1$.
Using eq.(\ref{vw}), we can write down the iterative relation between $P_1^n$ and $P_1^{n+1}$, 
\begin{equation}
    \begin{split}
        P_1^{n+1} 
        &= 
        {\left|\langle1|U_0^{n+1}|\psi(x)\rangle\right|}^2P_0^n
        +
        {\left|\langle1|U_1^{n+1}|\psi(x)\rangle\right|}^2P_1^n
        \\
        &=
        \cos^2(x - w_1^{n+1})P_0^n + \cos^2(x - v_1^{n+1})P_1^n
        \\
        &=
        [\frac{1}{2} + \frac{1}{2}\cos(2x - 2w_1^{n+1})](1-P_1^n)
        +
        [\frac{1}{2} + \frac{1}{2}\cos(2x - 2v_1^{n+1})]P_1^n
        \\
        &=
        \frac{1}{2} + \frac{1}{2}\cos(2x - 2w_1^{n+1})
        +
        \frac{1}{2}\left[
        \cos(2x - 2v_1^{n+1}) - \cos(2x - 2w_1^{n+1})
        \right]
        P_1^n
        \\
        &=
        A_{n+1}(x) + B_{n+1}(x)P_1^n
    \end{split}
\end{equation}
where $A_{n}(x)$ and $B_{n}(x)$ are defined as
\begin{equation}
    A_{n}(x) =  \frac{1}{2} + \frac{1}{2}\cos(2x - 2w_1^{n})
\end{equation}
\begin{equation}
    B_{n}(x) = 
    \frac{1}{2}\left[
    \cos(2x - 2v_1^{n+1}) - \cos(2x - 2w_1^{n+1})
    \right]
\end{equation}
Notice that $P_1^0 = A_0(x)$, we have
\begin{equation}
    P_n^1=
    \sum_{j=1}^{n}{
    \left[
    A_j(x)\prod_{k=j+1}^{n}B_k(x)
    \right]
    }
    \label{Pnbasic}
\end{equation}
$B_n(x)$ can be rewritten as
\begin{equation}
    \begin{split}
        B_{n}(x) &= 
        \frac{1}{2}\left[
        \cos(2x - 2v_1^{n+1}) - \cos(2x - 2w_1^{n+1})
        \right]
        \\
        =&
        \frac{1}{2}\left[
        \cos2x\cos2v_1^{n+1} + \sin2x\sin2v_1^{n+1} 
        - \cos2x\cos2w_1^{n+1} - \sin2x\sin2w_1^{n+1}
        \right]
        \\
        =&
         \frac{1}{2}\left[
        \cos2x(\cos2v_1^{n+1} - \cos2w_1^{n+1})
        + \sin2x(\sin2v_1^{n+1} - \sin2w_1^{n+1})
        \right]
        \\
        =&
        \frac{1}{2}
        \sqrt{(\cos2v_1^{n+1} - \cos2w_1^{n+1})^2 
        + (\sin2v_1^{n+1} - \sin2w_1^{n+1})^2}
        \\
        &\cdot\cos{\left[
        2x - \arctan2(\sin2v_1^{n+1} - \sin2w_1^{n+1}, \cos2v_1^{n+1} - \cos2w_1^{n+1})
        \right]}
        \\
        =&
        |\sin(v_1^n - w_1^n)|\cdot
        \cos{\left(
        2x - \beta_n
        \right)}
    \end{split}
    \label{Bn}
\end{equation}
where we denote $\beta_n = \arctan2(\sin2v_1^{n+1} - \sin2w_1^{n+1}, \cos2v_1^{n+1} - \cos2w_1^{n+1})$. Substitute eq.(\ref{Bn}) into eq.(\ref{Pnbasic}),
\begin{equation}
    \begin{split}
        P_n^1 =&
        \sum_{j=1}^{n}{
        \left\{
        [\frac{1}{2} + \frac{1}{2}\cos(2x - 2w_1^{n})]
        \prod_{k=j+1}^{n}
        |\sin(v_1^k - w_1^k)|\cdot
        \cos{\left(
        2x - \beta_k
        \right)}
        \right\}
        }
        \\
        =&
        \frac{1}{2} +
        \frac{1}{2}\cos(2x - 2w_1^{1})
        \prod_{k=2}^{n}
        |\sin(v_1^k - w_1^k)|\cdot
        \cos{\left(
        2x - \beta_k
        \right)}
        \\
        &+\frac{1}{2}
        \sum_{j=1}^{n}
        \left\{
        \left[
        |\sin(v_1^{j+1} - w_1^{j+1})| 
        \cos{\left(
        2x - \beta_{j+1}
        \right)}
        +
        \cos(2x-2w_1^{j+1})
        \right]
        \right.
        \\
        &\left.
        \cdot
        \prod_{k=j+2}^{n}
        |\sin(v_1^k - w_1^k)|\cdot
        \cos{\left(
        2x - \beta_k
        \right)}
        \right\}
    \end{split}
\end{equation}
Notice that
\begin{equation}
    \begin{split}
         &|\sin(v_1^{j+1} - w_1^{j+1})| 
         \cos{\left(
         2x - \beta_{j+1}
        \right)}
         +
        \cos(2x-2w_1^{j+1})
        \\
        =&
        \cos2x\cdot
        \left[
        |\sin(v_1^{j+1} - w_1^{j+1})|\cos\beta_{j+1} + \cos2w_1^{j+1}
        \right]
        \\
        &+
        \sin2x\cdot
        \left[
        |\sin(v_1^{j+1} - w_1^{j+1})|\sin\beta_{j+1} + \sin2w_1^{j+1}
        \right]
        \\
        =&
        \eta_n\cos(2x-\zeta_n)
    \end{split}
\end{equation}
where for simplicity, we denote
\begin{equation}
    \eta_n = \sqrt{1+\sin^2(v_1^{j+1} - w_1^{j+1})+|\sin(v_1^{j+1} - w_1^{j+1})|\cos(2w_1^{j+1}-\beta_{j+1})}
\end{equation}
\begin{equation}
    \zeta_n = 
    \arctan2(|\sin(v_1^{j+1} - w_1^{j+1})|\sin\beta_{j+1} + \sin2w_1^{j+1},
    |\sin(v_1^{j+1} - w_1^{j+1})|\cos\beta_{j+1} + \cos2w_1^{j+1})
\end{equation}
Then we can rewrite $P_n^1$ as
\begin{equation}
    \begin{split}
        P_n^1 
        =&
        \frac{1}{2} +
        \frac{1}{2}\cos(2x - 2w_1^{1})
        \prod_{k=2}^{n}
        |\sin(v_1^k - w_1^k)|\cdot
        \cos{\left(
        2x - \beta_k
        \right)}
        \\
        &+
        \frac{1}{2}\sum_{j=1}^n
        \left\{
        \eta_n\cos(2x-\zeta_n)
        \cdot
        \prod_{k=j+2}^{n}
        |\sin(v_1^k - w_1^k)|\cdot
        \cos{\left(
        2x - \beta_k
        \right)}
        \right\}
    \end{split}
\end{equation}

\begin{figure}
    \centering
    \includegraphics[width=0.85\linewidth]{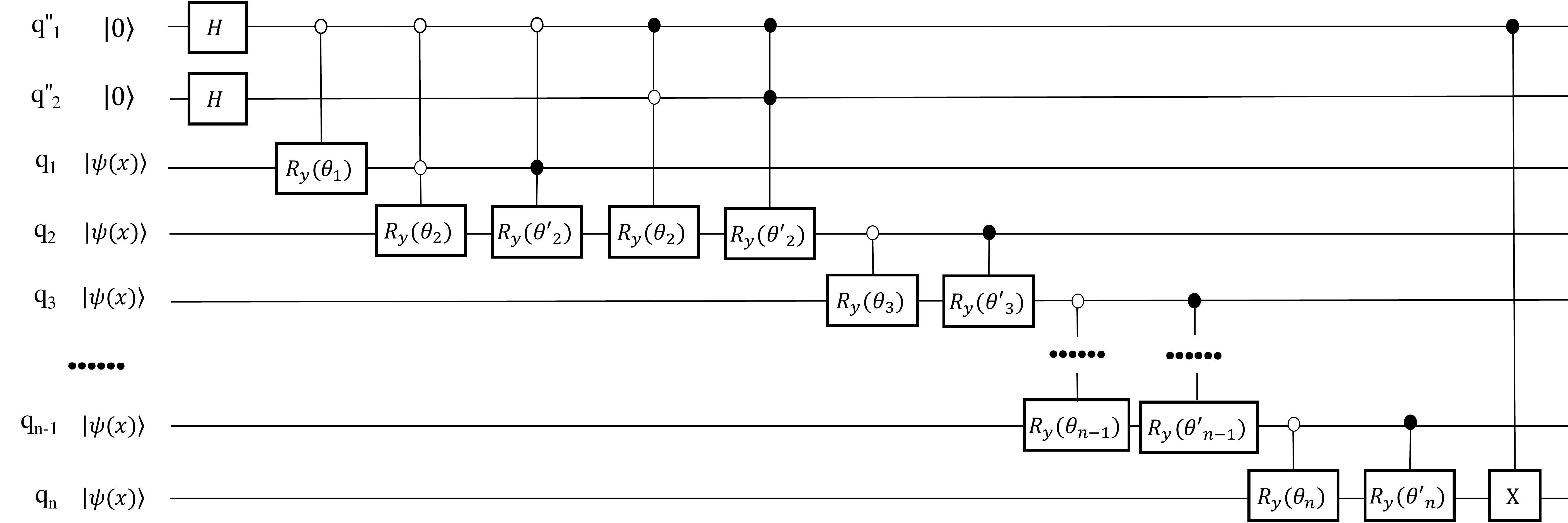}
    \caption{
    {\bf Sketch of quantum circuit $U_n({\bf\Theta_n})$.}
    \\
    $U_n(\Theta_n)$ is the basic modulus of the whole circuit that can estimate outputs with given function $F_N$.
    ${\bf\Theta_n}$ contains all parameters $\theta, \theta'$ or $v, w$ of the control rotation gates.
    $U_n(\Theta_n)$ can generate components of Fourier series no greater than the $n$-th order.
    Different from the simple qubit chain shown in fig.(\ref{figs_chain}), there are two auxiliary qubits $q''_1$ and $q''_2$ in quantum circuit $U_n(\Theta_n)$.
    $q''_1$ and $q''_2$ are initially prepared at state $|0\rangle$, and will acts as control gates after the Hadamard (H) gate.
    Mathematically, it can generates output as described in eq.(\ref{output_un}).
    In the main article, there is additionally a swap gate in $U_n$, which ensures all $U_n$ share the same qubit carrying outputs.
    }
    \label{figs_uj}
\end{figure}

Further, if we replace the initial state of $q_1$ with $|+\rangle$, discard $U^1$, and keep all the rest the same, then the probability to get $|1\rangle$ when measuring $q_N$ will be
\begin{equation}
    \begin{split}
        {P_n^1}'
        =&
        \frac{1}{2} +
        \frac{1}{2}\sum_{j=1}^n
        \left\{
        \eta_n\cos(2x-\zeta_n)
        \cdot
        \prod_{k=j+2}^{n}
        |\sin(v_1^k - w_1^k)|\cdot
        \cos{\left(
        2x - \beta_k
        \right)}
        \right\}
    \end{split}
\end{equation}
Hence, for the circuit shown in fig.(\ref{figs_uj}), if we apply $Z$ measurement on $q_N$ after the whole operation, the probability to get $|1\rangle$ will be
\begin{equation}
    \frac{1}{2}(P_n^1+1-{P_n^1}') = 
    \frac{1}{2}\cos(2x - 2w_1^{1})
    \prod_{k=2}^{n}
    |\sin(v_1^k - w_1^k)|\cdot
    \cos{\left(
    2x - \beta_k
    \right)}
    +\frac{1}{2}
    \label{output_un}
\end{equation}
In the main article we denote the circuit shown in fig.(\ref{figs_uj}) as $U_n({\bf \Theta_n})$, where there are totally $n$ qubits.
We use ${\bf \Theta_n}$ to represent parameters ${\theta}$ and ${\theta'}$ of the rotation gates.
Thus, at most $N$ operations are required to rebuild $F_N(x)$, denote them as $U_n, n=1,2,\cdots,N$.
$U_n|\psi(x)\rangle^{\otimes n}$ can generate Fourier series no greater than the $n$-th order.
$U_n$ contains a control operation chain that acts on $q''_1, q''_2, q_{N-n+1}, q_{N-n+2}, \cdots, q_n$, where $q''_1, q''_2$ are still auxiliary qubits and now the control operation chain starts from qubit $q_{N-n+1}$.
Generally in each operation $U_n$, there are several free variable sharing the similar notation, like $v, w, \beta$.
To avoid misleading notations, we introduce superscript to indicate the operation $U_n$, and define
\begin{equation}
    \alpha = \frac{1}{2}
    \prod_{k=2}^{n}
    |\sin(v_1^k - w_1^k)|
\end{equation}
when $k\geq2$
Thus, if we apply $U_n$ on $N$ input qubits all prepared at state $|\psi(x)\rangle$, the probability to get $|1\rangle$ when measuring the final qubit $q_N$ will be
\begin{equation}
    \begin{split}
        \sum_{j=1}^{2^{N-1}}
        {\left|
        {\left[
        |j\rangle_{(q_1, q_2, \cdots, q_{N-1})}
        \otimes|1\rangle
        \right]}^\dagger
        U_n|\psi(x)\rangle^{\otimes N}
        \right|}^2
        =
        \alpha^n
        \prod_{k=1}^{n}
        \cos{\left(
        2x - \beta^n_k
        \right)}
        +\frac{1}{2}
    \end{split}
    \label{prob_un}
\end{equation}
while we have $\alpha = 1/2$ and $\beta = 2w_1^1$ when $k=1$.

The present scheme can be extended to higher dimensions.
By replacing the input state from $|\psi(x)\rangle^{\otimes n}$ into $|\psi(x)\rangle^{\otimes n_x}\otimes|\psi(x)\rangle^{\otimes n_y}$, 
the probability to get $|1\rangle$ when measuring the final qubit $q_N$ after $U_n$ will be
\begin{equation}
    \alpha^n
        \prod_{k=1}^{n_x}
        \cos{\left(
        2x - \beta^n_k
        \right)}
        \prod_{k=n_x+1}^{n}
        \cos{\left(
        2y - \beta^n_k
        \right)}
        +\frac{1}{2}
\end{equation}
where $n_x, n_y\geq0$ and $n_x+n_y=n$.
With this approach multi variable can be included into the scheme, leading to the potential applications for higher dimensions.

To extend the method into complex varaiables, the real and complex terms must be calculated separately.  In most classical fast Fourier transform (FFT) algorithms, both the real terms and complex terms are included.  Therefore, at least two parallel quantum circuits are required if one would like to calculate the complex functions based on our algorithm.

\section{Design of $U_{pre}$}
\label{sec_upre}
\begin{figure}
    \centering
    \includegraphics[width=0.45\linewidth]{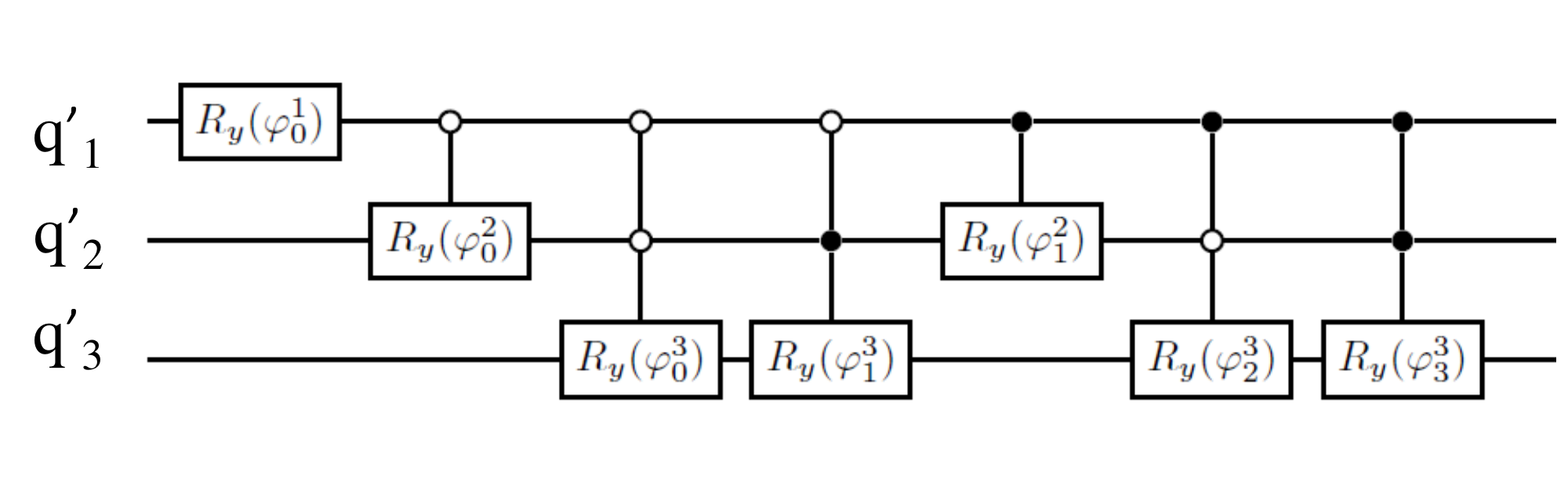}
    \caption{
    {\bf A sketch of $U_{pre}$ that acts on three qubits.}
    \\
    To convert the three qubits $q'_1,q'_2,q'_3$ into $|\psi'_f({\bf\gamma})\rangle$, the iterative process is repeated for three times.
    In the first iterative 'loop', we only need to apply a single rotation gate on $q'_1$.
    In the second loop, $q'_1$ performs as a control qubit.
    Two different rotation gates are applied on $q'_2$ corresponding to the two eigenstates of $q'_1$.
    Similarly, in the third loop, $q'_1$ and $q'_2$ together play the role of control qubits, and there are four rotation gates acting on $q'_3$.
    Totally, three are one rotation y gate acting on $q'_1$, two control rotation gates and four control control rotation gates.
    }
    \label{figs_ctrl}
\end{figure}

$U_{pre}$ is introduced to convert the auxiliary qubits $q'_1, q'_2, \cdots, q'_M$ into $|\psi'_f\rangle$
\begin{equation}
    |\psi'_f({\bf \gamma})\rangle =
    \sum_{n=0}^{2^M-1}
    \sqrt{\gamma_n}
    |n\rangle
\end{equation}
where ${\bf \gamma=(\gamma_0, \gamma_1, \cdots, \gamma_{2^M-1})}$, $\gamma_n\geq0$ and $\sum_{n=0}^{2^M-1}\gamma_n = 1$.
Here we provide a universal but not optimal design of $U_{pre}$. Denote operation $U_{pre}({\bf \gamma})$ that satisfies
\begin{equation}
    U_{pre}({\bf \gamma})|0\rangle = |\psi'_f({\bf \gamma})\rangle
\end{equation}
There is an iterative method to design $U_{pre}({\bf \gamma})$, when there are more than 2 components($M>1$) in ${\bf\gamma}$,
\begin{equation}
    U_{pre}({\bf \gamma})=
    R_y(2\omega)\otimes
    \left[
    |0\rangle\langle0|\otimes
    U_{pre}({\bf \gamma_A}) 
    +
    |1\rangle\langle1|
    U_{pre}({\bf \gamma_B})
    \right]
\end{equation}
Otherwise if ${\bf \gamma}$ only contains 2 terms($M=1$),
\begin{equation}
    U_{pre}({\bf \gamma})=R_y(2\omega)
\end{equation}

where $\omega=\arctan2(\sum_{n=2^{M-1}}^{2^M-1}\gamma_n, \sum_{n=0}^{2^{M-1}-1}\gamma_n)$,
and as long as $0<\sum_{n=2^{M-1}}^{2^M-1}\gamma_n<1$, we define
\begin{equation}
    {\bf \gamma_A} = \frac{1}{\sum_{n=0}^{2^{M-1}-1}\gamma_n}
    (\gamma_0, \gamma_1, \cdots, \gamma_{2^{M-1}-1})
\end{equation}
\begin{equation}
    {\bf \gamma_B} = \frac{1}{\sum_{n=2^{M-1}}^{2^M-1}\gamma_n}
    (\gamma_{2^{M-1}}, \gamma_{2^{M-1}+1}, \cdots, \gamma_{2^{M}-1})
\end{equation}
Here we set the first half of original ${\bf\gamma}$ as $\gamma_A$ and the second half as $\gamma_B$.
Therefore $\gamma_A$ and $\gamma_B$ is cut off from $\gamma_{2^{M-1}-1}$.
If we rewrite $n$ into binary numbers, the $n-$th component $\gamma_n$ will be assigned into $\gamma_A$ when $n=(0X\cdots X)_2$, while it will be assigned into $\gamma_B$ when $n=(1X\cdots X)_2$, subscripts denote binary description.
As for the exception, if $\sum_{n=2^{M-1}}^{2^M-1}\gamma_n=0$, we can set all components in ${\bf \gamma_B}$ as 0, and same to the situation when $\sum_{n=0}^{2^{M-1}-1}\gamma_n=1$.
As a simple but not trivial example, fig.(\ref{figs_ctrl}) is a design of  $U_{pre}$ that acts on three qubits $q'_1,q'_2,q'_3$, which totally contains one rotation y gate acting on $q'_1$, two control rotation gates and four control control rotation gates.

Then consider constraints of $\gamma.$
If we measure the last qubit $q_N$ after the whole operation, probability to get $|1\rangle$ is
\begin{equation}
    \begin{split}
        &\sum_{m=1}^{2^M}
        \sum_{n=1}^{2^{N-1}}
        {\left|
        \left[
        |m\rangle_{q'_1,\cdots,q'_M}
        \otimes|n\rangle_{q_1,\cdots,q_{N-1}}\otimes
        |1\rangle_{q_N}
        \right]^\dagger
        U_{pre}\otimes U_f
        |0\rangle^{\otimes M}\otimes|\psi(x)\rangle^{\otimes N}
        \right|}^2
        \\
        =&
        \sum_{m=1}^{2^M}
        \sum_{n=1}^{2^{N-1}}
        {\left|
        \left[
        |m\rangle_{q'_1,\cdots,q'_M}
        \otimes|n\rangle_{q_1,\cdots,q_{N-1}}\otimes
        |1\rangle_{q_N}
        \right]^\dagger
        \left[
        \sum_{j=0}^{2^M-1}|j\rangle\langle j|\otimes U_j({\bf \Theta_j})
        \right]
        |\phi_f\rangle\otimes|\psi(x)\rangle^{\otimes N}
        \right|}^2
        \\
        =&
        \sum_{m=1}^{2^M}
        \sum_{n=1}^{2^{N-1}}
        {\left| 
        \sum_{j=0}^{2^M-1}
        \langle m|j\rangle_{q'_1,\cdots,q'_M}
        \left[
        |n\rangle_{q_1,\cdots,q_{N-1}}\otimes
        |1\rangle_{q_N}
        \right]^\dagger
        \cdot
        \langle j|\phi_f\rangle
        U_j({\bf \Theta_j})
        |\psi(x)\rangle^{\otimes N}
        \right|}^2
        \\
        =&
        \sum_{m=1}^{2^M}
        \sum_{n=1}^{2^{N-1}}
        {\left| 
        \sum_{j=0}^{2^M-1}
        \delta_{mj}
        \left[
        |n\rangle_{q_1,\cdots,q_{N-1}}\otimes
        |1\rangle_{q_N}
        \right]^\dagger
        \cdot
        \langle j|\phi_f\rangle
        U_j({\bf \Theta_j})
        |\psi(x)\rangle^{\otimes N}
        \right|}^2
        \\
        =&
        \sum_{m=1}^{2^M}
        \sum_{n=1}^{2^{N-1}}
        {\left| 
        \left[
        |n\rangle_{q_1,\cdots,q_{N-1}}\otimes
        |1\rangle_{q_N}
        \right]^\dagger
        \cdot
        \langle m|\phi_f\rangle
        U_m({\bf \Theta_m})
        |\psi(x)\rangle^{\otimes N}
        \right|}^2
        \\
        =&
        \sum_{m=1}^{2^M}
        \sum_{n=1}^{2^{N-1}}
        {\left| 
        \sqrt{\gamma_m}
        \left[
        |n\rangle_{q_1,\cdots,q_{N-1}}\otimes
        |1\rangle_{q_N}
        \right]^\dagger
        U_m({\bf \Theta_m})
        |\psi(x)\rangle^{\otimes N}
        \right|}^2
    \end{split}
\end{equation}
Substitute eq.(\ref{prob_un}), we can continue the above deduction,
\begin{equation}
    \begin{split}
        &\sum_{m=1}^{2^M}
        \sum_{n=1}^{2^{N-1}}
        {\left|
        \left[
        |m\rangle_{q'_1,\cdots,q'_M}
        \otimes|n\rangle_{q_1,\cdots,q_{N-1}}\otimes
        |1\rangle_{q_N}
        \right]^\dagger
        U_{pre}\otimes U_f
        |0\rangle^{\otimes M}\otimes|\psi(x)\rangle^{\otimes N}
        \right|}^2
        \\
        =&
        \sum_{m=1}^{2^M}
        \sum_{n=1}^{2^{N-1}}
        {\left| 
        \sqrt{\gamma_m}
        \left[
        |n\rangle_{q_1,\cdots,q_{N-1}}\otimes
        |1\rangle_{q_N}
        \right]^\dagger
        U_m({\bf \Theta_m})
        |\psi(x)\rangle^{\otimes N}
        \right|}^2
        \\
        =&
        \sum_{m=1}^{2^M}
        \gamma_m
        \sum_{n=1}^{2^{N-1}}
        {\left| 
        \left[
        |n\rangle_{q_1,\cdots,q_{N-1}}\otimes
        |1\rangle_{q_N}
        \right]^\dagger
        U_m({\bf \Theta_m})
        |\psi(x)\rangle^{\otimes N}
        \right|}^2
        \\
        =&
        \sum_{m=1}^{2^M}
        \gamma_m
        \left[
        \alpha^m
        \prod_{k=1}^{m}\cos(2x-\beta_k^m)
        +\frac{1}{2}
        \right]
        \\
        =&
        \sum_{m=1}^{N}
        \gamma_m
        \alpha^m
        \prod_{k=1}^{m}\cos(2x-\beta_k^m)
        +\frac{1}{2}
    \end{split}
\end{equation}

Subsequent to the whole operation, the output state is
\begin{equation}
    \begin{split}
    |\Psi_{out}\rangle
    =
    \sqrt{-CF_N(x_j)+\frac{1}{2}}|\phi_0\rangle_{q',q'',q_1,\cdots,q_{N-1}}
    \otimes|0\rangle_{q_N}
    +\sqrt{CF_N(x_j)+\frac{1}{2}}|\phi_1\rangle_{q',q'',q_1,\cdots,q_{N-1}}
    \otimes|1\rangle_{q_N}
    \end{split}
    \label{output}
\end{equation}
where $|\phi_{0, 1}\rangle$ describing output state of all qubits except the last one $q_N$, and $C$ is a constant ensuring $\frac{1}{2}\pm CF_N(x_j)\geq0$.
Hence, information stored in $q_N$ is essential and sufficient.
After measuring $q_N$, the frequency getting $|0\rangle$ is an estimation of $F_N(x)$.
Whereas, it is also appropriate to apply succeeding operations on $q_N$, regarding it as an intermediate.

To estimate $F_N(x)$, we need to ensure that
\begin{equation}
    C\sum_{n=1}^Na_n\cos(2nx+b_n)+\frac{1}{2}
    =
    \sum_{m=1}^{N}
    \gamma_m
    \alpha^m
    \prod_{k=1}^{m}\cos(2x-\beta_k^m)+\frac{1}{2}
    \label{constraint}
\end{equation}
where $\gamma$ gives $|\psi'_f\rangle$.

Theoretically, any design that satisfies eq.(\ref{constraint}) can be used to estimate $F_N(x)$.
Here we only give one solution set that we used in our simulation. Denote $a_n', b_n'$ as
\begin{equation}
    C\sum_{n=1}^{j}a'_n\cos(2nx+b'_n)
    =
    \sum_{n=1}^Na_n\cos(2nx+b_n)
    -\sum_{m=j+1}^{N}
    \gamma_m
    \alpha^m
    \prod_{k=1}^{m}\cos(2x-\beta_k^m)
    ,
    \qquad
    1\leq j<N
\end{equation}
Then for $n>1$ we can set
\begin{equation}
    \left\{
    \begin{array}{lr}
        \gamma_n = \frac{2^{n-1}Ca'_n}{\alpha^n} &
        \\
        \beta_k^n = -\frac{b'_n}{n}, & b'_n\neq0
        \\
        \beta_k^n = \frac{\pi}{3}\delta_{10}-\frac{\pi}{3(n-1)}, &b'_n=0
    \end{array}
    \right.
    \label{betagamma}
\end{equation}

\section{Implementation on IBM QASM simulator}

\begin{figure}[h]
    \centering
    \includegraphics[width=0.95\linewidth]{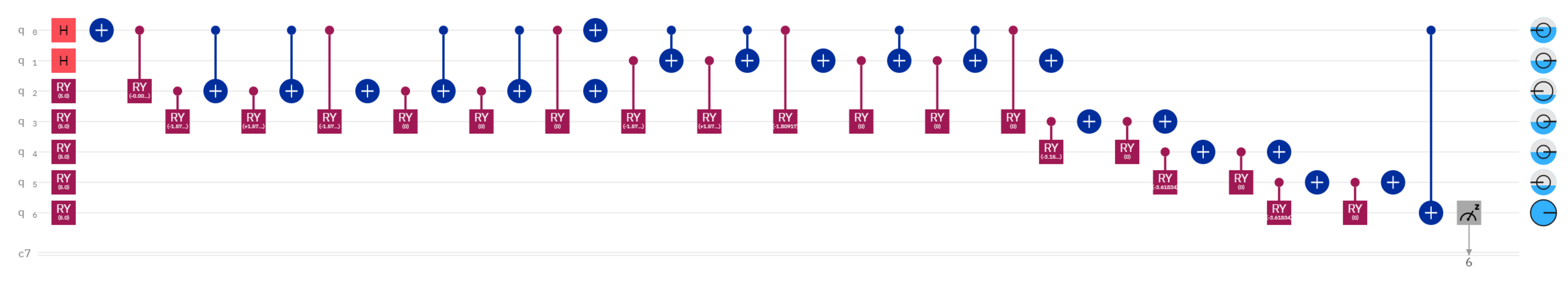}
    \caption{
    {\bf A sketch of $U_{5}$ implemented on IBM QASM simulator.}
    \\
    There are two auxiliary qubits $q_{0,1}$ and five qubits $q_2, \cdots, q_6$ carrying the input information $x$.
    For simplicity, other qubits are not included.
    The control-control-RY gates are decomposed into two CNOT gates and three control-RY gates.
    }
    \label{figs_u5}
\end{figure}

In the main article, we implemented $U_{3,5,7}$ independently on  IBM QASM simulator.
Fig.(\ref{figs_u5}) is a sketch of the implementation of $U_5$, where there are two auxiliary qubits $q_{0,1}$ and five qubits $q_2, \cdots, q_6$ carrying the input information $x$.
The control-control-RY gates are decomposed into two CNOT gates and three control-RY gates.
$U_5$ is designed to generate the component $cos^5(2x-0.0046)$.
Notice that the corresponding $\gamma_5$ is negative, we set,
\begin{equation}
\left\{
    \begin{array}{lr}
        \theta_n = 0\qquad, n>1
        \\
        \theta'_n = -2\cdot[(\beta_n-\pi/2)\mod\pi]
        \qquad, n>1
        \\
        \theta'_0=\theta_0 = -\beta_0
    \end{array}
    \right.
\end{equation}
where $\beta_n=0.0046$.
Measure the last qubit $q_6$ after the whole operation, probability to get state 0, namely $P_0$, will be proportional to $cos^5(2x-0.0046)$.
In other words, by this design we generate the component $-cos^5(2x-0.0046)$ in the final output, as $P_1$ is used for the estimation of $F_N(x)$.
To implement $U_n$, totally $O(n)$ CNOT gates and control-RY gates and NOT gates are required.
On the other hand, these setting can also be used for a positive term together with a final NOT gate applying on the last qubit.

Generally there are three approaches when optimizing the parameters with the constraint shown in Eq.(7).
One approach is to optimize the parameters ensuring that $|C|$ is as large as possible.
The larger  the $|C|$ value is, the less measurements is required to obtain the estimation.
Another approach is to optimize the parameters ensuring that the circuit is as shallow as possible.
The last approach is to make a balance, ensuring that the total time consuming is as less as possible.
Here we choose the second approach, ensuring that the circuit is as shallow as possible.
In this example, as the square wave function is an odd function, our design with $\gamma_{2,4,6}=0$ reduces half $U_{n}$ operations.
The special choice of $\theta$ and $\theta'$ ensure that half of them are zero, so that we can further reduce it to about  half the gates.  We need to stress that the our setting is not optimal, especially when one prefer a larger  value of $|C|$ instead of a shallow circuit.